\documentclass[sigconf]{acmart}
\usepackage{xcolor}
\usepackage{colortbl}
\usepackage{multirow}
\usepackage{subcaption}
\usepackage{tabularx}
\usepackage{enumitem}
\usepackage{bm}
\usepackage{xspace}

\newcommand{\sect}[1]{Section~\ref{#1}}

\newcommand{\fig}[1]{Figure~\ref{#1}}

\newcommand{\tab}[1]{Table~\ref{#1}}

\makeatletter
\DeclareRobustCommand\onedot{\futurelet\@let@token\@onedot}
\def\@onedot{\ifx\@let@token.\else.\null\fi\xspace}

\def\eg{\emph{e.g}\onedot} 
\def\ie{\emph{i.e}\onedot}

\def\etal{\emph{et al}\onedot}
\makeatother

\definecolor{mydarkblue}{rgb}{0,0.08,1}
\definecolor{mydarkgreen}{rgb}{0.02,0.6,0.02}
\definecolor{mydarkred}{rgb}{0.8,0.02,0.02}
\definecolor{mydarkorange}{rgb}{0.40,0.2,0.02}
\definecolor{mypurple}{rgb}{111,0,255}
\definecolor{myred}{rgb}{1.0,0.0,0.0}
\definecolor{mygold}{rgb}{0.75,0.6,0.12}
\definecolor{myblue}{rgb}{0,0.2,0.8}
\definecolor{mydarkgray}{rgb}{0.66,0.66,0.66}

\definecolor{lightbluegreen}{HTML}{C6E9ED}
\definecolor{lightblue}{HTML}{89BEE5}

\newcommand{\mybfparagraph}[1]{\noindentparagraph{\textbf{#1}}}

\newcommand{\mybfitvparagraph}[1]{\noindentparagraph{\textbf{#1}}}
\newcommand{\bfparagraph}[1]{\noindentparagraph{\textbf{#1}}}

\def\x{$\times$\xspace}

\def\sota{state-of-the-art\xspace}

\def\arch{PointAcc\xspace}
\def\archsmall{PointAcc.Edge\xspace}

\def\pointnetpp{PointNet++\xspace}

\def\minkunet{MinkowskiUNet\xspace}

\def\SparseConv{SparseConv\xspace}
\def\sparsemapping{mapping\xspace}
\def\mappingpair{map\xspace}
\def\mappingpairs{maps\xspace}

\def\knn{\textit{k}-nearest-neighbors\xspace}
\def\kernelmapping{kernel mapping\xspace}
\def\fps{farthest point sampling\xspace}
\def\ballquery{ball query\xspace}

\def\SparseMapping{Mapping\xspace}
\def\KernelMapping{Kernel Mapping\xspace}
\def\FPS{Farthest Point Sampling\xspace}
\def\KNN{\textit{k}-Nearest-Neighbors\xspace}
\def\BallQuery{Ball Query\xspace}

\def\ComparatorStruct{ComparatorStruct\xspace}
\def\MappingUnit{Mapping Unit\xspace}
\def\MatrixComputingUnit{Matrix Unit\xspace}
\def\MemoryManagementUnit{Memory Management Unit\xspace}
\def\streaming{fetch-on-demand\xspace}
\def\Streaming{Fetch-On-Demand\xspace}
\def\MPU{MPU\xspace}
\def\MMU{MMU\xspace}
\def\MCU{MXU\xspace}
\def\MIR{MIR\xspace}
\def\MIRContainer{MIR Container\xspace}

\def\mesorasi{Mesorasi\xspace}

\def\mesorasiAcc{9.1\%\xspace}

\AtBeginDocument{%
  \providecommand\BibTeX{{%
    \normalfont B\kern-0.5em{\scshape i\kern-0.25em b}\kern-0.8em\TeX}}}

\setcopyright{acmcopyright}
\copyrightyear{2021}
\acmYear{2021}
\setcopyright{rightsretained}
\acmConference[MICRO '21]{MICRO-54: 54th Annual IEEE/ACM International Symposium on Microarchitecture}{October 18--22, 2021}{Athens, Greece}
\acmBooktitle{MICRO-54: 54th Annual IEEE/ACM International Symposium on Microarchitecture (MICRO '21), October 18--22, 2021, Virtual Event, Greece}
\acmDOI{10.1145/3466752.3480084}
\acmISBN{978-1-4503-8557-2/21/10}

\begin{document}

\title{\arch: Efficient Point Cloud Accelerator}

\author{Yujun Lin, Zhekai Zhang, Haotian Tang, Hanrui Wang, Song Han}
\affiliation{%
  \institution{MIT}
  \city{Cambridge}
  \state{\{yujunlin, zhangzk, kentang, hanrui, songhan\}@mit.edu}
  \country{USA}
  \postcode{02139}
}
\email{https://pointacc.mit.edu}


\renewcommand{\shortauthors}{Lin, et al.}

\begin{abstract}
Deep learning on point clouds plays a vital role in a wide range of applications such as autonomous driving and AR/VR. These applications interact with people in real time on edge devices and thus require low latency and low energy. 
Compared to projecting the point cloud to 2D space, directly processing 3D point cloud yields higher accuracy and lower \#MACs.
However, 
the extremely sparse nature of point cloud
poses challenges to hardware acceleration.
For example, we need to explicitly determine the nonzero outputs and search for the nonzero neighbors (mapping operation),
which is unsupported in existing accelerators.
Furthermore, explicit gather and scatter of sparse features are required, resulting in large data movement overhead. 

In this paper, we comprehensively analyze the performance bottleneck of modern point cloud networks on CPU/GPU/TPU.
To address the challenges, we then present \arch, a novel point cloud deep learning accelerator.
\arch maps diverse mapping operations onto one versatile ranking-based kernel,
streams the sparse computation with configurable caching, and temporally fuses consecutive dense layers to reduce the memory footprint.
Evaluated on 8 point cloud models across 4 applications, \arch achieves 3.7\x speedup and 22\x energy savings over RTX 2080Ti GPU. Co-designed with light-weight neural networks, \arch rivals the prior accelerator Mesorasi by 100\x speedup with \mesorasiAcc higher accuracy running segmentation on the S3DIS dataset. \arch paves the way for efficient point cloud recognition.
\end{abstract}

\ccsdesc[500]{Computer systems organization~Neural networks}
\keywords{point cloud, neural network accelerator, sparse convolution}


\maketitle

\begin{figure}
    \centering
    \includegraphics[width=\linewidth]{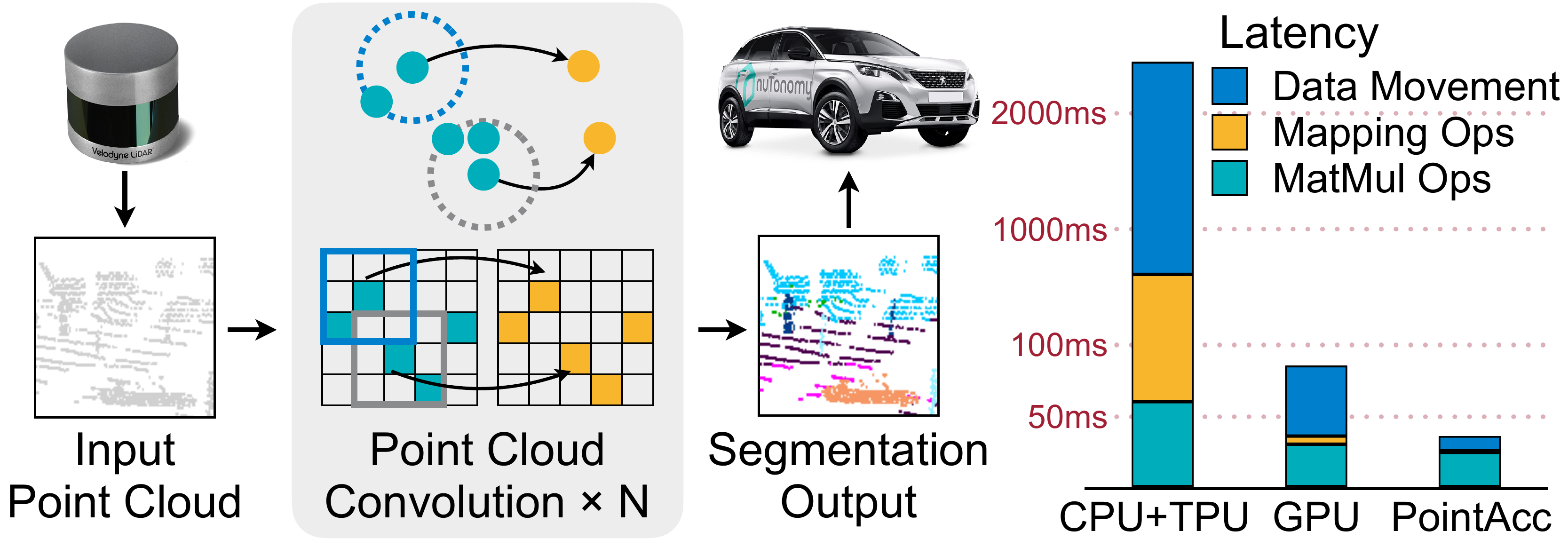}
    \caption{Point cloud deep learning is crucial for real-time AI applications. PointAcc accelerates point cloud computations by resolving sparsity and data movement bottlenecks.}
    \label{fig:teaser}
\end{figure}
\begin{figure}[t]
    \centering
    \includegraphics[width=\linewidth]{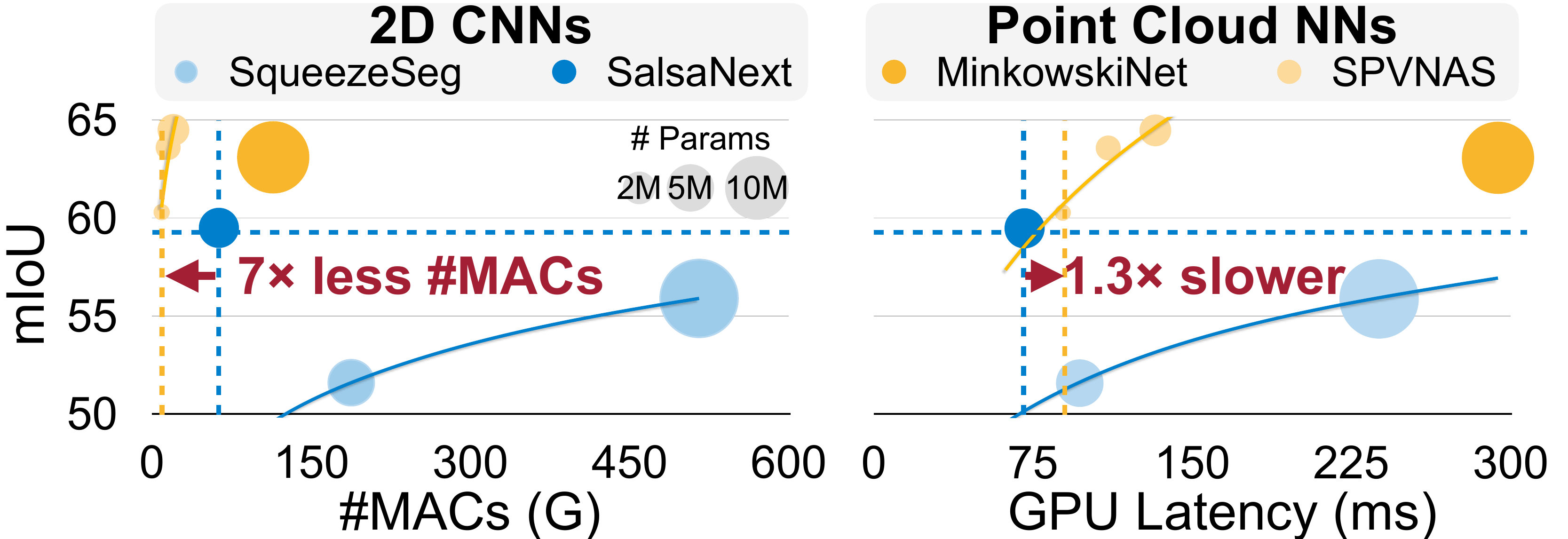}
    \caption{Compared to 2D CNNs, point cloud networks have higher accuracy and lower \#MACs, but higher GPU latency due to low utilization brought by sparsity and irregularity.}
    \label{fig:teaser_2dvs3d}
\end{figure}
\section{Introduction}
\begin{figure*}[t]
    \centering
    \includegraphics[width=\linewidth]{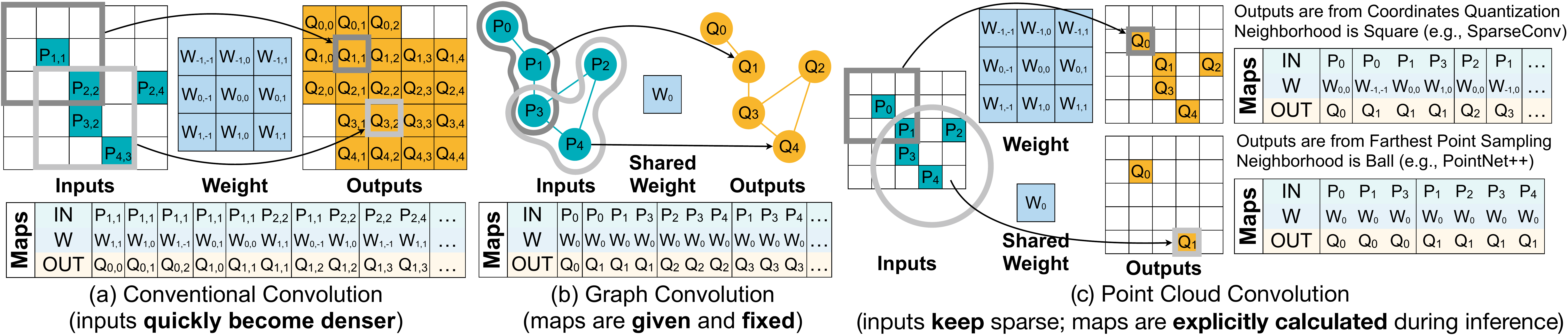}
    \caption{Convolution on point clouds (c) is very different from convolution on images (a) and graphs (b).}
    \label{fig:conv}
\end{figure*}

A point cloud is a collection of points that represent a physical object or 3D scene. Point clouds are usually generated by sensors like LiDARs at a rapid speed (2 million points per second for a 64-channel LiDAR sensor). 
As LiDARs are becoming as cheap as just hundreds of dollars, they are extensively deployed everywhere, in cars, robots, drones, and even in iPhone 12 Pros. Consequently, point clouds have become a modality as important as images and videos for deep learning applications such as autonomous driving, photography, virtual reality (VR), and augmented reality (AR). 
These applications require real-time interactions with humans, and thus it is crucial to emphasize not only high accuracy, but also low latency and energy consumption.

Compared to projecting 3D point cloud to 2D then applying Convolution Neural Networks (CNN) on 2D flattened point clouds (\fig{fig:teaser_2dvs3d} (left)), directly processing 3D point clouds with Point Cloud Networks~\cite{qi2017pointnet++, choy20194d, li2018pointcnn, wu2019pointconv, wang2018dgcnn, li2021deepgcns, xu2018spidercnn, graham20183d, tang2020searching} yields up to 5\% higher accuracy with 7\x less \#MACs. 
However, point cloud networks run significantly slower on existing general-purpose hardware than CNNs (\fig{fig:teaser_2dvs3d} right). The \sota point cloud model \minkunet (114G MACs) runs at only 11.7 FPS even on a powerful NVIDIA RTX 2080Ti GPU, while ResNet50~\cite{he2015deep} (4G MACs) with similar input size can run at 840 FPS. To the best of our knowledge, the only accelerator for point cloud networks so far is Mesorasi~\cite{feng2020mesorasi}. However, the ``delayed aggregation'' technique used by Mesorasi is restricted to only a small fraction of point cloud models, where all the neighbors are restricted to share the same weights. In contrast, \sota point cloud networks~\cite{choy20194d, tang2020searching} use different weights for different neighbors, offering much higher accuracy.
To tackle such dilemma, we present \arch, an efficient domain-specific accelerator for point cloud deep learning.

Computing on point clouds is challenging due to the high sparsity nature of inputs. For instance, outdoor LiDAR point clouds usually have a density of less than 0.01\%, while traditional CNNs take in 100\% dense images. 
Moreover, the sparsity in point clouds is fundamentally different from that in traditional CNNs which comes from the weight pruning and ReLU activation function. The sparsity in point clouds conveys physical information: the sparsity pattern is constrained by the physical objects in the real world. That is to say, the nonzero points will never dilate during the computation.

Therefore, point cloud processing requires a variety of \textit{mapping operations}, such as \ballquery and \kernelmapping, to establish the relationship between input and output points for computation, which has not been explored by existing deep learning accelerators. To tackle this challenge, \arch unifies these operations in a ranking-based computation paradigm which can generalize to other similar operations. By leveraging this shared computation paradigm, \arch further presents a versatile design to support diverse mapping operations on the arbitrary scale of point clouds.
Moreover, strictly restricted sparsity pattern in point cloud networks
leads to irregular sparse computation pattern. Thus it requires explicit gather and scatter of point features for matrix computation, which results in a massive memory footprint. To address this, \arch performs flexible control on on-chip memory using decoupled and explicit data orchestration~\cite{pellauer2019buffets}. By caching the input features on demand with configurable block size and temporally fusing the computation of consecutive layers, \arch manages to improve the data reuse and reduce the expensive DRAM access.

In summary, this work makes the following contributions:
\begin{itemize}
    \item We comprehensively investigate the datasets, computation cost and memory footprint of point cloud processing, and analyze the performance bottleneck on various hardware platforms. The sparsity of point cloud introduces unexplored mapping operations, and requires explicit gather/scatter.
    
    \item We present a versatile design to support diverse mapping operations that finds the nonzero neighbors and nonzero output point clouds corresponding to each weight. It unifies and converts the mapping operations into ranking-based comparisons. Our design manages to handle the arbitrary scales of point clouds.

    \item We present an efficient memory management design that decouples the memory request and response to precisely control the memory. It exploits caching and simplifies the layer fusion for point cloud networks, which reduces the DRAM access by up to 6.3\x.
\end{itemize}
We implement and synthesize \arch in TSMC 40nm technology node. Extensively evaluated with 8 modern point cloud networks on 5 datasets, \arch achieves an order of magnitude speedup on average compared with other hardware platforms. Co-designing the neural network, \arch outperforms the prior \sota point cloud accelerator Mesorasi by 100\x speedup and \mesorasiAcc better mIoU accuracy running segmentation on the S3DIS dataset.

\section{Background}
\label{sect:background}

\begin{figure*}[t]
    \centering
    \includegraphics[width=\linewidth]{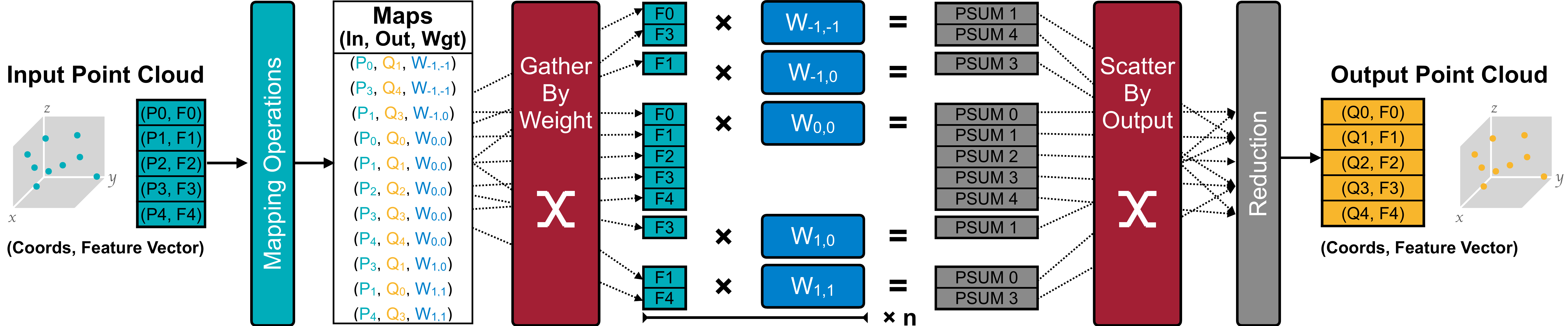}
    \caption{Existing CPU / GPU implementations for a point cloud convolution layer. Here $\bm{p}_k = (x_k, y_k, z_k)$.}
    \label{fig:block-structure}
\end{figure*}
\begin{table*}[htbp]
  \centering
  \caption{Point Cloud Convolutions in Point Cloud Deep Learning.}
  \newcolumntype{g}{>{\columncolor{lightbluegreen}}c}
  \newcolumntype{e}{>{\columncolor{lightblue}}c}
  \scalebox{1.0}{
  \begin{tabular}{c|cc|c}
    \toprule
    \multirow{2}[2]{*}{\textbf{Point Cloud Convolution}} & \multicolumn{2}{g|}{\textbf{Mapping Operations}} & \multicolumn{1}{e}{\textbf{MatMul Operations}} \\
     & \textbf{Output Cloud Construction} & \textbf{Neighbor Search}  & \textbf{Neighbor Aggregation} \\
    \midrule
    PointNet++-based & Farthest Point Sampling\cite{qi2017pointnet++,li2018pointcnn} & Ball Query~\cite{qi2017pointnet++} &  MaxPool~\cite{qi2017pointnet++,wang2018dgcnn,li2021deepgcns} \\
    (including Graph-based) &  Random Sampling~\cite{li2018pointcnn}    & \textit{k} Nearest Neighbors\cite{li2018pointcnn,wang2018dgcnn,li2021deepgcns,he2020svga}  & Convolution1d~\cite{li2018pointcnn} \\
    \midrule
    SparseConv-based & \multirow{2}[2]{*}{Coordinates Quantization} & \multirow{2}[2]{*}{Kernel Mapping} & Accumulation \\
    \cite{graham20183d,choy20194d,tang2020searching,han2020occuseg,shi2021pvrcnn++,cheng2021af2s3net,yin2021center} &       &      &  (\ie, Convolution3d) \\
    \bottomrule
  \end{tabular}
  }
  \label{tab:block-structure}
\end{table*}

\emph{Point Cloud} is a set of points $\bm{x} = \{\bm{x}_k\} = \{(\bm{p}_k, \bm{f}_k)\}$, where $\bm{p}_k = (x_k, y_k, z_k)$ is the coordinate of the $k$\textsuperscript{th} point, and $\bm{f}_k$ is the corresponding 1-D feature vector.
The key component of point cloud networks is the point cloud convolution. We compare convolution on different modalities in \fig{fig:conv}, where the green grids/dots are (nonzero) inputs and the yellow ones are outputs. 

Similar to image convolution which works on the receptive field (\fig{fig:conv}a), point cloud convolution is conducted on the \emph{neighborhood} of the output point (\fig{fig:conv}c). Intuitively, if a input point $\bm{p}_j$ is the $n$-th neighbor of output point $\bm{q}_k$, we will perform $\bm{psum}_k$+=$\bm{f}_j\bm{w}_n$, where $\bm{w}_n$ is the corresponding weights. We define such relationship between input and output point as a \textbf{\emph{\mappingpair}}, \ie, \mappingpair is a tuple  $(\bm{p}_{j}, \bm{q}_{k}, \bm{w}_{n})$.
Point cloud convolution iterates over all \mappingpairs and performs multiplication-accumulation accordingly.
Note that \mappingpairs in image convolution can be directly inferred by pointer arithmetic since image pixels are dense (\fig{fig:conv}a), and \mappingpairs in graph convolution (\fig{fig:conv}b) are provided as the adjacency matrix and stay constant across layers. However, \mappingpairs in point cloud convolution have to be explicitly calculated \emph{every time} downsampling the point cloud, due to the sparse nature of point clouds. In addition, for different neighbors, graph convolutions use the same weights, while \sota point cloud convolutions use different weights.

The \sota CPU/ GPU implementation of point cloud convolution is summarized in \fig{fig:block-structure}. Specifically, we first perform \sparsemapping operations to find the input-output \mappingpairs. Based on these \mappingpairs, we gather the input features for different weights, transform features via matrix multiplication, and then scatter-aggregate the partial sums to the corresponding output points.
The entire computation process consists of three types of operations: \sparsemapping, data movement (gather/scatter) and matrix multiplication (MatMul), as summarized in \tab{tab:block-structure}. We categorize point cloud convolutions into two classes: PointNet++-based and SparseConv-based convolutions.
Graph-based convolutions~\cite{wang2018dgcnn,li2021deepgcns,he2020svga} are treated as the special case of PointNet++-based convolution, where the mapping operations work on the point features instead of point coordinates.

\subsection{\SparseMapping Operations}
\label{sect:background-mapping-operations}
\SparseMapping is the procedure to find the input-output \emph{\mappingpairs} $\mathcal{K}$ in point clouds, where $\mathcal{K} = \{(\bm{p}_{j}, \bm{q}_{k}, \bm{w}_{n})\}$. The search for \mappingpairs runs two tasks: output point cloud construction and neighbor search (\tab{tab:block-structure}). These operations usually only take point coordinates as input.

\subsubsection{Output Point Cloud Construction}
The coordinates of output points are explicitly calculated during downsampling. Upsampling the point cloud is the inverse of corresponding downsampling.

\begin{figure*}[t]
    \centering
    \begin{minipage}[b]{0.6\linewidth}
    \includegraphics[width=0.325\linewidth]{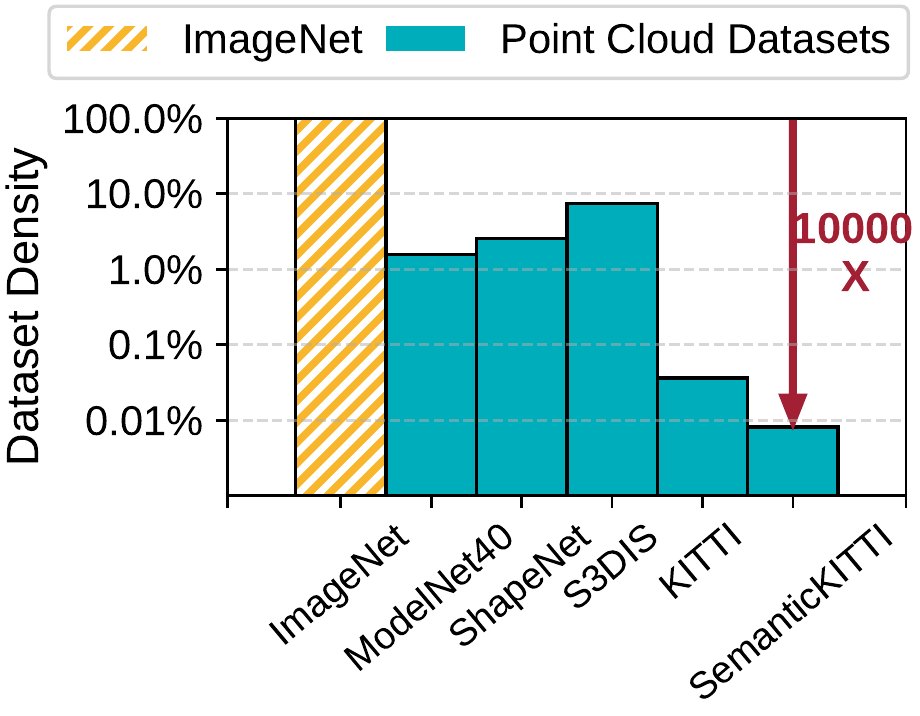}
    \includegraphics[width=0.325\linewidth]{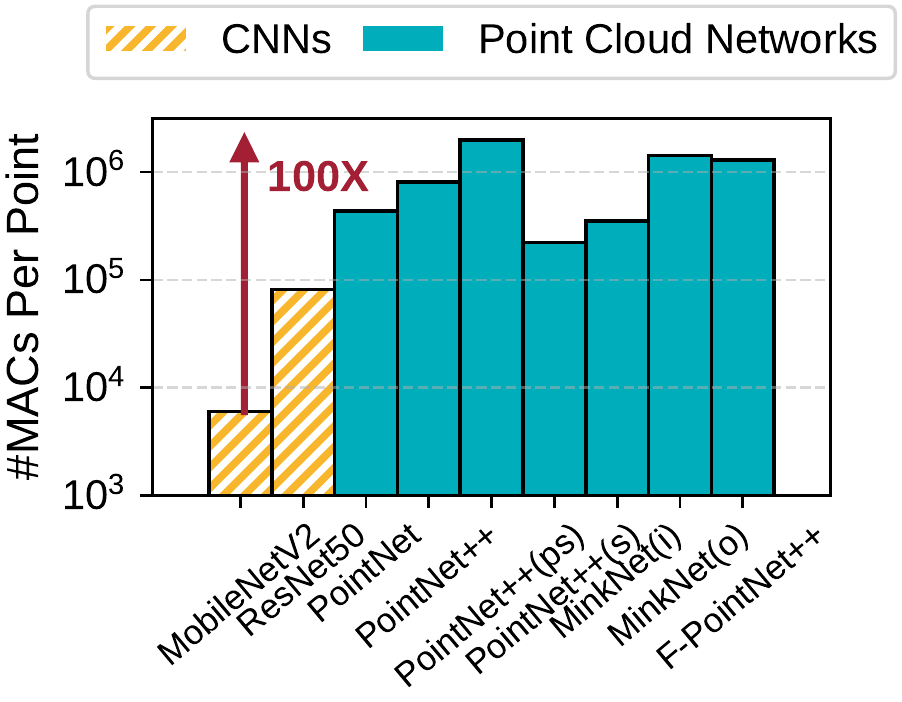}
    \includegraphics[width=0.325\linewidth]{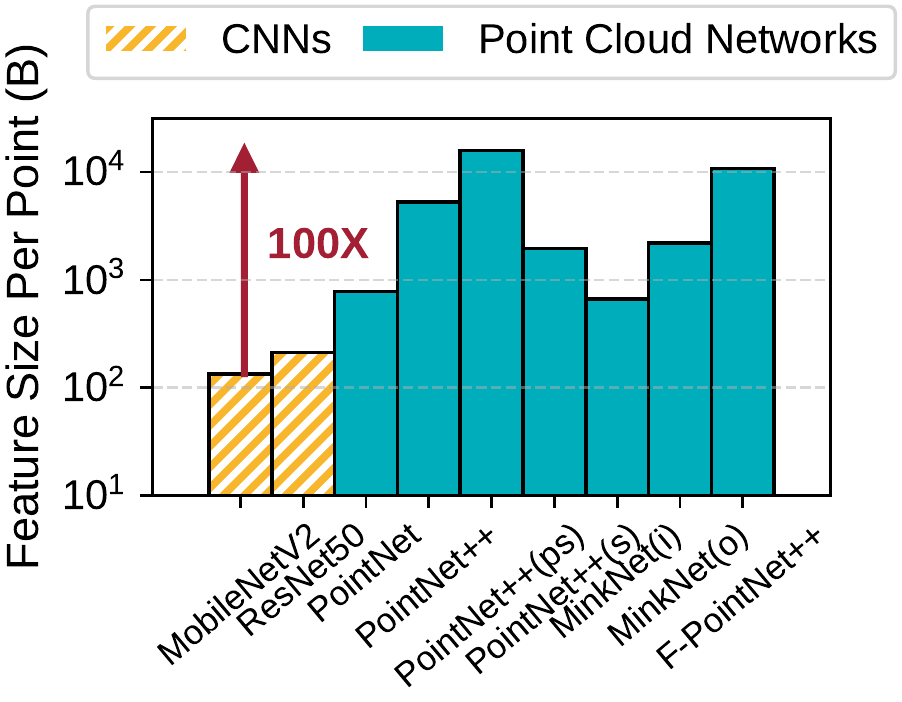}
    \caption{Point cloud datasets are ultra sparse. Point cloud networks have large computation cost and memory footprint.}
    \label{fig:pointnet-statistics}
    \end{minipage}
    \begin{minipage}[b]{0.38\linewidth}
    \includegraphics[width=\linewidth]{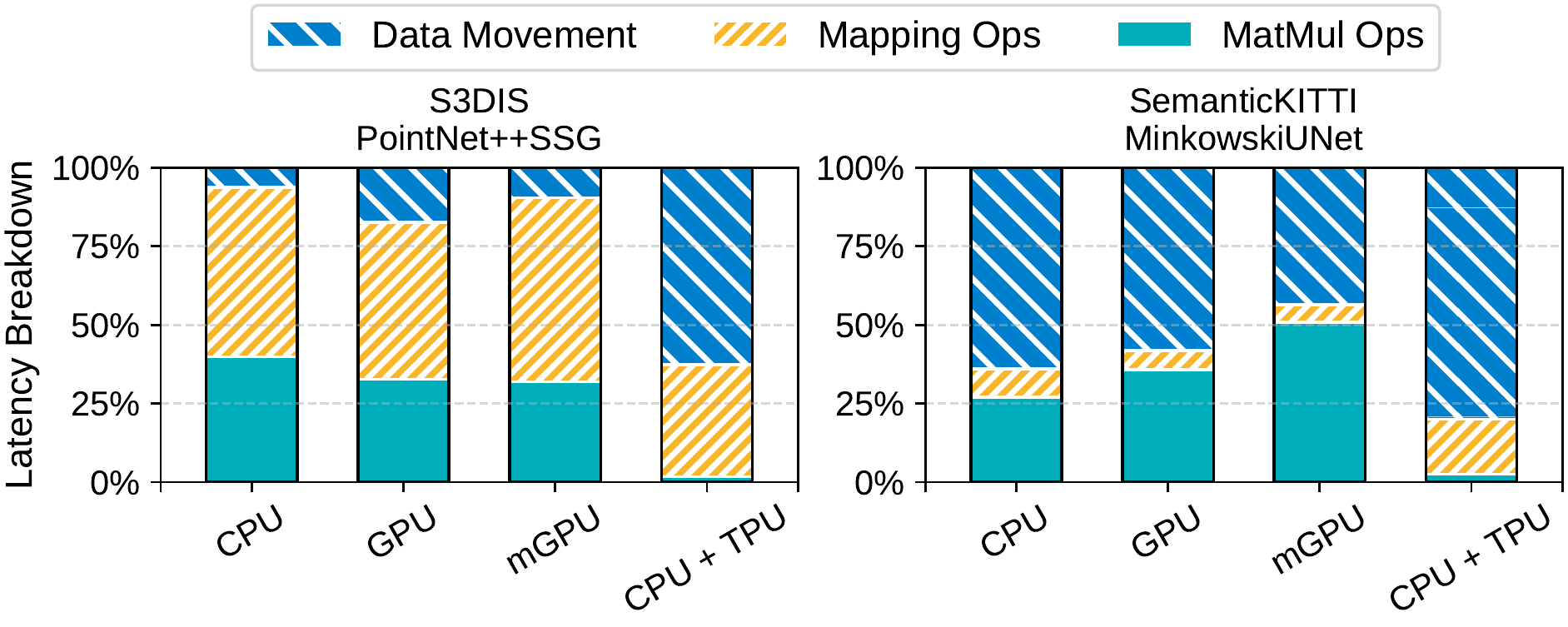}
    \caption{Point cloud networks are bottlenecked by data movement and mapping operations.}
    \label{fig:bottleneck-network}
    \end{minipage}
\end{figure*}

\mybfitvparagraph{Coordinates Quantization.}
\SparseConv-based convolution directly reduces the point cloud resolution during the downsampling. Specifically, the output point coordinate is calculated by quantization: $\bm{q} = \texttt{floor}(\bm{p} / ts) \times ts$, where $ts$ is the tensor stride ($ts = 2^k$ after $k$ downsamplings). For example, point $(3, 5)$ whose $ts=1$ will be quantized to $(2, 4)$ whose $ts=2$ after downsampling, and point $(4, 8)$ whose $ts=4$ will be quantized to $(0, 8)$ whose $ts=8$ after downsampling. 
Such quantization can be easily implemented on hardware by clearing the lowest $\log_{2}(ts)$ bits of coordinates.

\mybfitvparagraph{Farthest Point Sampling.}
\pointnetpp-based convolution applies \fps during the downsampling, where each output point is sampled from the input point cloud $I$ one by one iteratively. In each iteration $t$, we choose the point that has the largest distance to the current output point cloud $O_t$.
For example, in \fig{fig:conv}c (bottom), we select $\bm{q}_0$ as the first output point, and since $\bm{q}_4$ is farthest from $\bm{q}_0$, we select it as the second output point.

\subsubsection{Neighbor Search}
\label{sect:background-neighbor-search}
For each output point, the neighbor search is performed in the input point cloud to generate the \mappingpairs.

\mybfitvparagraph{Kernel Mapping.}
In \SparseConv-based convolution, each output point will travel through all its neighborhood positions with offsets $\bm{\delta}\in\{-1, 0, 1\}^D$, where $D$ is dimension of the point cloud ($D = 2$ in \fig{fig:conv}). In \fig{fig:conv}c, output point $\bm{q}_0$ has neighbor $\bm{p}_0$ with offset $(0, 0)$ and neighbor $\bm{p}_1$ with offset $(1, 1)$. Hence, \mappingpairs $(\bm{p}_0, \bm{q}_0, \bm{w}_{0, 0}), (\bm{p}_1, \bm{q}_0, \bm{w}_{1,1})$, are generated.

\mybfitvparagraph{\KNN and \BallQuery.}
In \pointnetpp-based convolution, based on the distances to output point $\bm{q}$, top-\textit{k} input points $\bm{p}$ are selected. Ball query further requires these points to lie in the sphere of radius $r$ , \ie, $||\bm{p} - \bm{q} ||_2 \leq r$.
In \fig{fig:conv}c (bottom), there are three \mappingpairs associated with $\bm{q}_0$, and four \mappingpairs for $\bm{q}_1$.

\subsection{MatMul Operations}
\label{sect:background-matmul-operations}
MatMul operations are conducted on features $\bm{f}$ based on the \mappingpairs. Specifically, we group all input features associated with the same weight $\bm{w}_n$ (\emph{i.e.} gather by weight) and use one~\cite{choy20194d} or several FC layers~\cite{qi2017pointnet++} to obtain partial sums $\bm{psum} = \bm{f}\cdot \bm{w}_{n}$. The partial sums are later aggregated (via max-pooling~\cite{qi2017pointnet++}, convolution~\cite{li2018pointcnn} or accumulation~\cite{choy20194d}) after being scattered to the corresponding output location (\emph{i.e.} scatter by output). In \fig{fig:block-structure}, we gather $[\bm{f}_{0}, \bm{f}_{3}]$, multiply them with the weight matrix $\bm{w}_{-1,-1}$, and scatter-aggregate them to output $[\bm{f}_{1}, \bm{f}_{4}]$ according to the \mappingpairs. We then repeat the same process for $\bm{w}_{-1,0}, ..., \bm{w}_{1,1}$ sequentially.
\section{Motivation}
\label{sect:motivation}

As shown in \fig{fig:pointnet-statistics} and \fig{fig:bottleneck-network}, we systematically profile the characteristics of point cloud datasets and networks, and their performance on various platforms, including CPU, GPU, mobile GPU, and TPU. We find that the challenge of accelerating point cloud networks comes from the intrinsic sparsity of the point cloud.

\bfparagraph{Challenge: High Input Sparsity.}
Unlike the input of image CNNs which are dense tensors, point cloud is naturally sparse. \fig{fig:pointnet-statistics} (left) plots the input sparsity of five mainstream point cloud datasets ModelNet40~\cite{wu2015modelnet}, ShapeNet~\cite{chang2015shapenet}, KITTI~\cite{geiger2012kitti}, S3DIS~\cite{armeni2016s3dis} and SemanticKITTI~\cite{behley2019semantickitti}, with details summarized in \tab{tab:benchmarks}. Point clouds of 3D indoor scenes and objects have a density of $<10^{-2}$, and the 3D outdoor scenes are even sparser, reaching a density of less than $10^{-4}$. In contrast, ImageNet has 100\% density at the input and 50\% density on average after ReLU, which is up to four orders of magnitude denser than point cloud.

Conventionally, the input sparsity in CNNs results from the ReLU activation function. On the contrary, the sparsity in point cloud networks comes from the spatial distribution of the points, which contains the physical information in the real world. Therefore, it places hard constraint on the sparsity pattern in the point cloud convolution. In traditional sparse CNNs, the nonzero inputs are multiplied with every nonzero weights, and thus the nonzeros will dilate in the output. Such regular computation pattern is exploited in the prior sparse NN accelerators~\cite{han2016eie,parashar2017scnn,albericio2016cnvlutin,zhang2016cambricon}. However, in point cloud NNs, each nonzero input point is not always multiplied with all nonzero weights: the relationship among input points, weights and output points are determined by \sparsemapping operations. Hence previous sparse NN accelerators will not work.

\bfparagraph{Bottleneck I. New Operations: \SparseMapping Operations.}
The first bottleneck due to point cloud sparsity is the \sparsemapping operations introduced in \sect{sect:background-mapping-operations}. As shown in \fig{fig:bottleneck-network} (left), the PointNet++-based networks spend more than 50\% of total runtime on \sparsemapping operations on general-purpose hardware platforms. 
Unfortunately, existing specialized NN accelerators do not support these \sparsemapping operations, and will worsen the performance.
We take TPU~\cite{jouppi2017tpu} as an example. TPUs are tailored for dense matrix multiplications. Therefore, we have to first move all relevant data to host memory, rely on CPU to calculate the \sparsemapping operations and gather the features accordingly, and then send back the contiguous matrices to the TPU. Such a round trip between the heterogeneous memory can be extremely time-consuming. In practice, we found that the data movement time takes up 60\% to 90\% of total runtime.

\bfparagraph{Bottleneck II. Large Memory Footprint.}
The second bottleneck resulting from point cloud sparsity is the large memory footprint. Since the point cloud convolution has to explicitly gather input features and scatter partial sums, features can be repeatedly accessed for at most $3^3$=27 times (3D kernel with the size of 3). Moreover, in \SparseConv-based models, downsampling only reduces the spatial resolution, and the number of points is usually not scaled down by 4$\times$ as in 2D CNNs.
Therefore, the memory footprint of features in point cloud networks significantly surpasses CNNs. As shown in \fig{fig:pointnet-statistics} (right), the memory footprint of the features per point in point cloud networks can achieve up to 16 KB, which is 100\x higher than CNNs. Thus the data movement alone can take up over 50\% of total runtime on CPUs and GPUs, as shown in \fig{fig:bottleneck-network} (right).

\begin{figure*}[t]
\begin{minipage}[b]{0.65\linewidth}
    \centering
    \includegraphics[width=\linewidth]{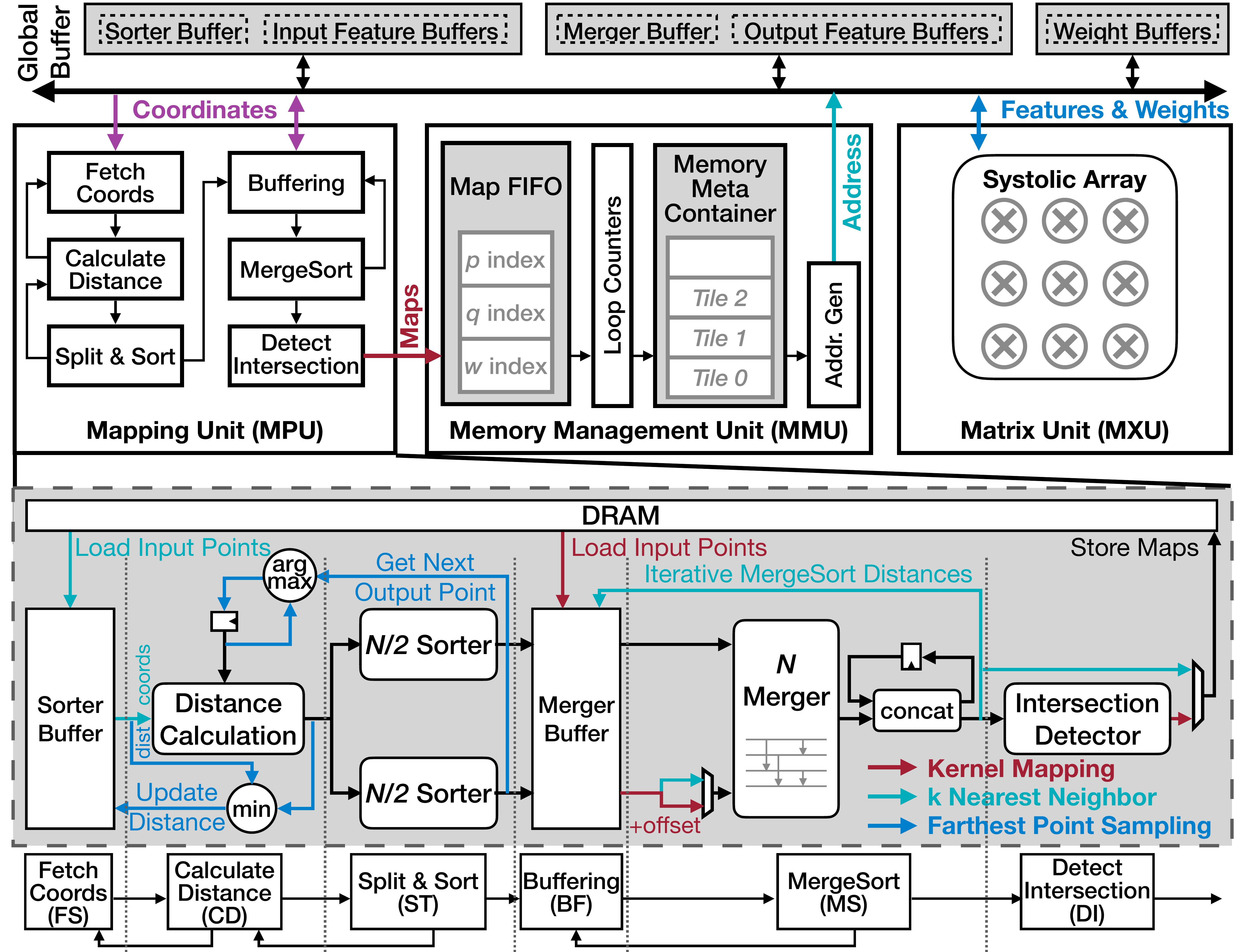}
    \caption{Overview of \arch Architecture.}
    \label{fig:architecture}
\end{minipage}
\hfill
\begin{minipage}[b]{0.345\linewidth}
    \centering
    \includegraphics[width=0.9\linewidth]{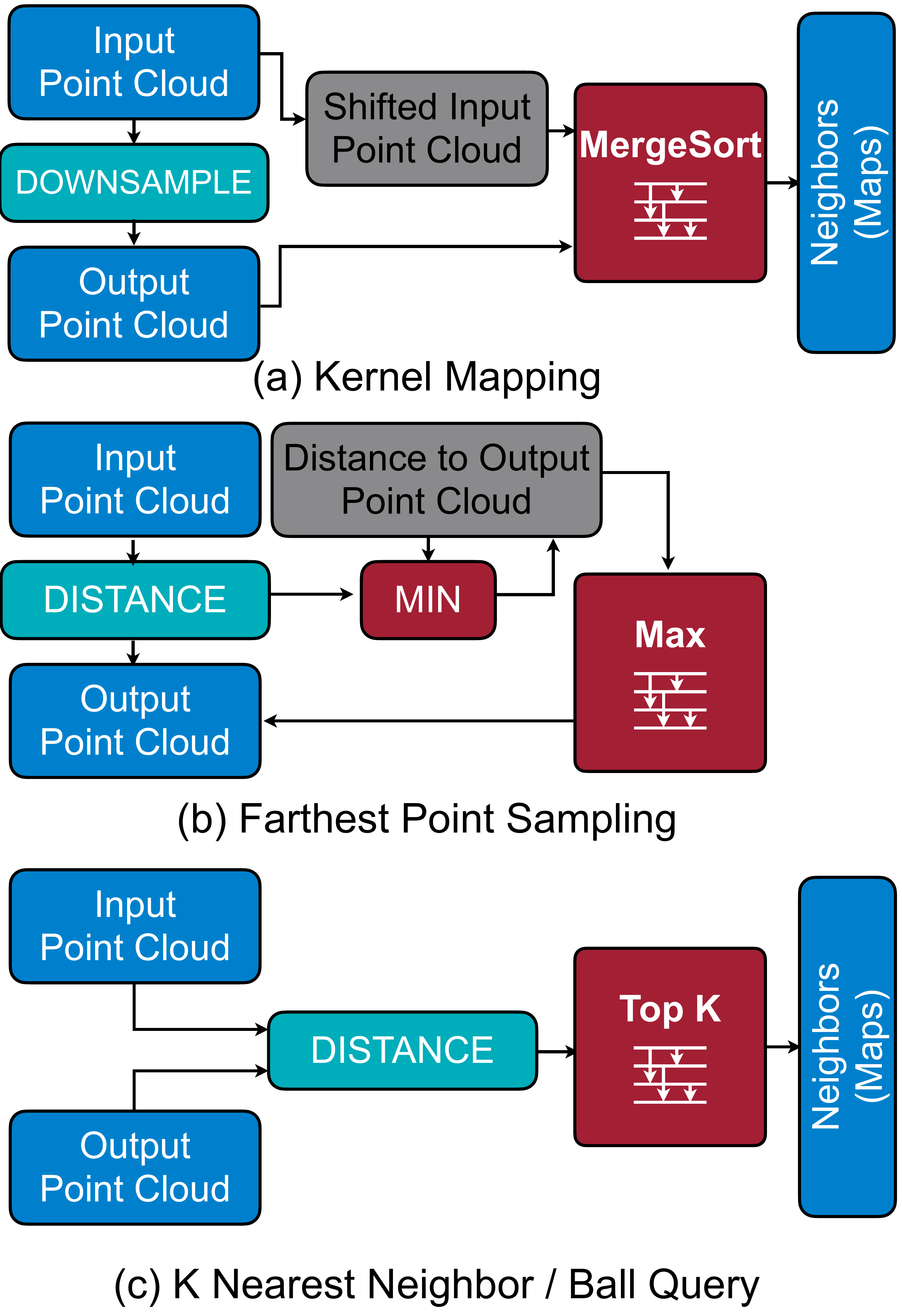}
    \caption{Mapping operations introduced by point cloud are unified to similar paradigm with ranking-based compute kernel.}
    \label{fig:neighbor-search-diagram}
\end{minipage}
\end{figure*}
\begin{figure}[t]
    \centering
    \includegraphics[width=\linewidth]{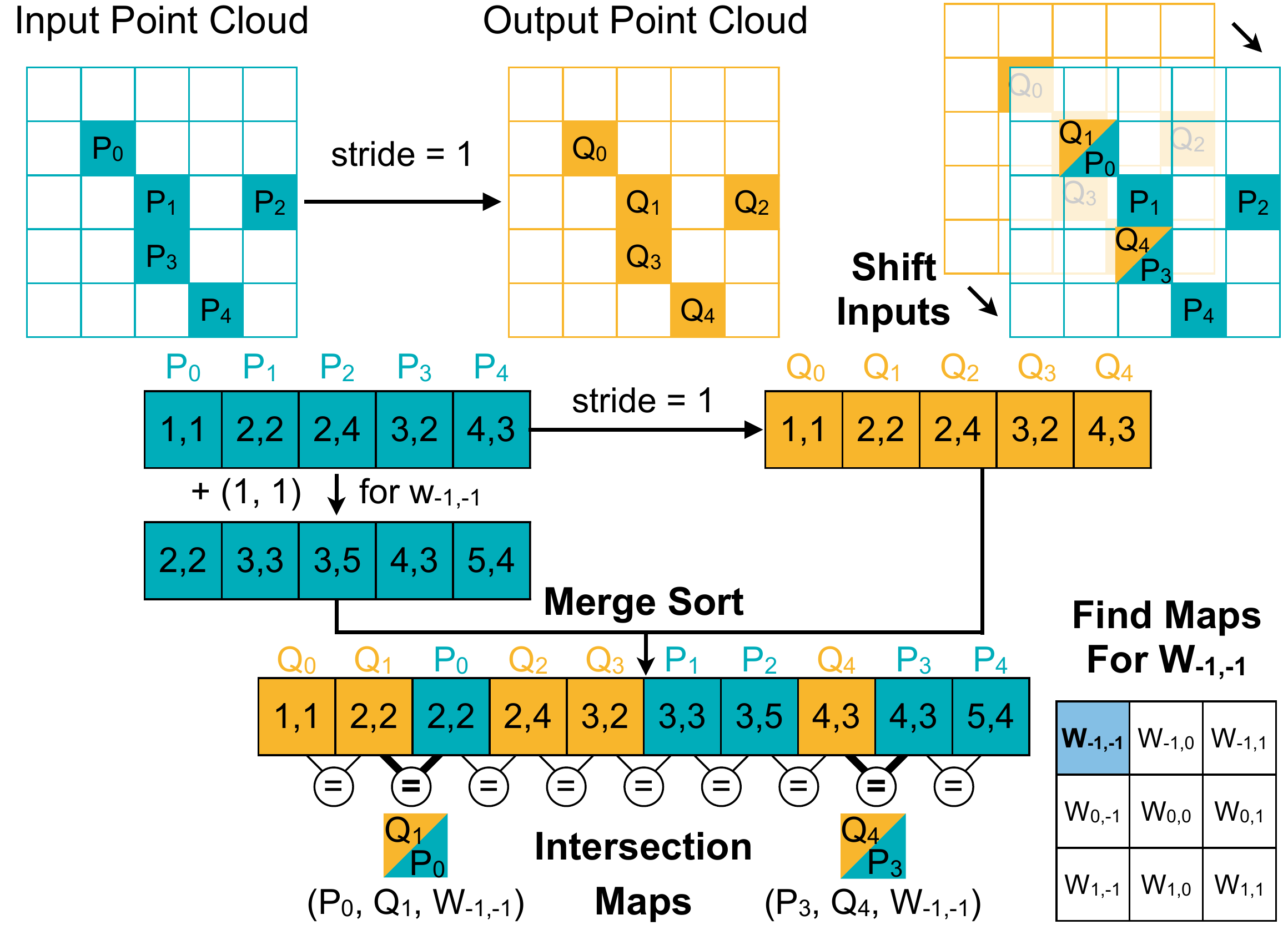}
    \caption{Mergesort-based Kernel Mapping Implementation.}
    \label{fig:spconv-example-alg}
\end{figure}

\begin{figure*}[t]
    \centering
    \includegraphics[width=\linewidth]{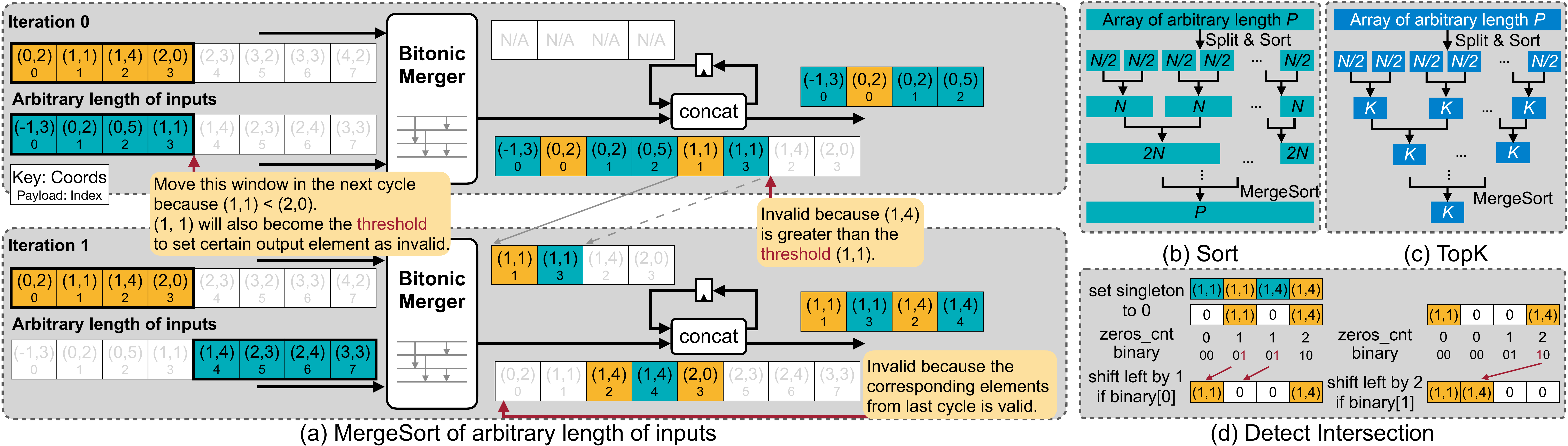}
    \caption{(a) \MappingUnit handles \texttt{MergeSort} of arbitrary length of inputs with forwarding loop; (b) \MappingUnit supports \texttt{Sort} of arbitrary length of inputs by iteratively MergeSort splited and sorted subarrays in a tree structure; (c) \MappingUnit flexibly realizes \texttt{TopK} on arbitrary length of inputs by truncating the intermediate merge-sorted subarrays to the length of \textit{k}; (d) The intersection detector taking \textit{N} elements has $\log N$ stages.}
    \label{fig:mpu-example}
\end{figure*}

\section{Architecture}
\label{sect:architecture}

To tackle the challenge discussed in \sect{sect:motivation}, we present \arch architecture design, as shown in \fig{fig:architecture}. It consists of three parts: \MappingUnit, \MemoryManagementUnit and \MatrixComputingUnit. \MemoryManagementUnit bridges the other two units by preparing the data for \MatrixComputingUnit based on the output of \MappingUnit.
By configuring the data flow in each unit, \arch flexibly supports various point cloud networks efficiently.

\subsection{\MappingUnit}
Conventional point cloud accelerators~\cite{gieseke2014buffer, qiu2009gpu, heinzle2008hardware, winterstein2013fpga, xu2019tigris, feng2020mesorasi} only focus on \knn which is only one of \sparsemapping operations in the domain. Our \MappingUnit (\MPU) targets all diverse \sparsemapping operations, including \kernelmapping, \knn, \ballquery and \fps.
Instead of designing specialized modules for each possible operations, we propose to unify these diverse \sparsemapping operations into one computation paradigm, and convert them to point-cloud-agnostic operations, which can be further used to design one versatile specialized architecture.

\subsubsection{Diverse \SparseMapping Ops in One Architecture}
\label{sect:unify-mapping-ops}

The ultimate goal of \sparsemapping operations is to generate \mappingpairs in the form of (input point index, output point index, weight index) tuple for further computation (\eg, convolution). We observe that no matter which algorithm is used, these \mappingpairs are always constructed based on the \textit{comparison} among distances. Thus we offer to convert these comparisons into ranking operations (\fig{fig:neighbor-search-diagram}), and process in parallel for different points in the point cloud (point-level parallelism).

\mybfparagraph{\FPS} obtains the point in the input point cloud with the largest distance to the current output point cloud. We simply convert it to a \texttt{Max} operation on distances (\fig{fig:neighbor-search-diagram}b). 

\mybfparagraph{\KNN} searches \textit{k} points in the input point cloud with smallest distances to the given output point, and \textbf{\ballquery} further requires these distances to be smaller than a predefined value. They can be implemented with \texttt{TopK} operation (\fig{fig:neighbor-search-diagram}c). 

\mybfparagraph{\KernelMapping} can be regarded as finding the input points with the exact distance in the specific direction to the output points, \ie, finding the input point $p$ in the input point cloud $I=\{\bm{p}\}$ with certain offset $\delta$ to the output cloud $O=\{\bm{q}\}$, \ie $\bm{p}=\bm{q}+\bm{\delta}$. For example, for the maps associating with weight $w_{-1, 0}$, the input points are all right above the output points with distance of 1. Hence, the comparison is \texttt{Equal} operation on distances, which is done via the hash table in the \sota implementation~\cite{tang2020searching}. The hash table records the input point cloud coordinates, and each output point will query its possible neighbor from the table. A query hit indicates a \mappingpair is founded. However, such hash-table-based solution is inefficient in terms of circuit specialization. On one hand, we cannot afford a large on-chip SRAM for the hash table which could be as large as 160 MB considering the input point cloud size and the load factor. On the other hand, even if we exploit the locality in the query process and build the hash table on the fly, we could not parallelize it efficiently. A parallelized hash table requires random parallelized read to the SRAM, which typically requires an $N$-by-$N$ crossbar with a space complexity of $O(N^2)$.

Instead, we model the \kernelmapping as finding the \textit{intersection} on point coordinates between output point cloud $O$ and shifted input point cloud $I' = \{\bm{p} - \bm{\delta} | \bm{p} \in I\}$. The parallelized \texttt{Equal} operation between two point clouds can be further converted to \texttt{MergeSort} of two point clouds (\fig{fig:neighbor-search-diagram}c). The input point cloud $I$ is first applied the offset $-\bm{\delta}$, and then merge-sorted with the output point cloud $O$, which is an optionally downsampled version of the input cloud. The intersection can then be easily found by examining the adjacent elements, keeping those with the same coordinate and removing others.
Experiment shows that our mergesort-based solution could provide 1.4\x speedup while saving up to 14\x area compared to the hash-table-based design with the same parallelism.

\mybfitvparagraph{Example.} \fig{fig:spconv-example-alg} illustrates an example where the input point cloud (green) is multiplied with the weight $w_{-1,-1}$. $\bm{\delta} = (-1, -1)$ for $w_{-1,-1}$; thus the input point cloud is shifted in the $(1, 1)$ direction, \ie, the right-bottom direction.
The shifting is performed by adding each coordinate with $-\bm{\delta}=(1,1)$. For example, $p_{0}=(1,1)$ becomes $(2,2)$ and $p_{1}=(2,2)$ becomes $(3,3)$.
The shifted input cloud (green) are then merge-sorted with the output point cloud (yellow) which has the same coordinates as the input cloud since stride $s=1$, forming one sorted array. Each pair of adjacent elements in the array are fed to a comparator to see if their coordinates are equal.
For instance, shifted $p_{0}$ and $q_1$ shares the same coordinates $(2,2)$, and thus they form a \mappingpair $(p_0, q_1, w_{-1,-1})$. Here we found 2 \mappingpairs.

\subsubsection{\MappingUnit Architecture Design}
As all \sparsemapping operations are converted to point-cloud-agnostic ranking operations (\eg, MergeSort, TopK in \fig{fig:neighbor-search-diagram}),  \MPU exploits sorting-network-based design to support these ranking operations, and eliminates the data movement between co-processors as in TPU case in \fig{fig:bottleneck-network}.

We denote the comparator input element as \ComparatorStruct which contains the comparator key (coordinates or distance) and the payload (\eg, the point index).
As in \fig{fig:architecture}, \MPU has 6 stages:
\begin{itemize}[topsep=0pt, leftmargin=18pt]
\item \textbf{FetchCoords (FS)}: fetch \ComparatorStruct from sorter buffer; write back the updated distances forwarded from stage CD when running \fps (blue).

\item \textbf{CalculateDistance (CD)}: calculate the distances from input points to a specific output point; compare these distances with recorded distance in the payload and forward the minimum to the previous stage FS for \fps (blue).

\item \textbf{Sort (ST)}: split the outputs of stage CD into two sub-arrays, and sort them independently; compare the present maximum of sorter outputs with the history maximum in the register, and forward the final maximum to the previous stage CD after the traversal of the whole point cloud when executing \fps (blue).

\item \textbf{Buffering (BF)}: buffer the sorted arrays of \ComparatorStruct from the previous stage ST or from the later stage MS when running \knn (green).

\item \textbf{MergeSort (MS)}: merge-sort two arrays into one array (a forwarding loop is inserted inside the merger to handle arbitrary length of inputs); forward the results to the previous stage BF for sorting the arbitrary length of inputs when running \knn (green). 

\item \textbf{DetectIntersection (DI)}: detect the duplicate elements in the merged array as in \fig{fig:mpu-example}d. This stage is bypassed unless running \kernelmapping (red).

\end{itemize}
\begin{figure*}[t]
    \centering
    \includegraphics[width=\linewidth]{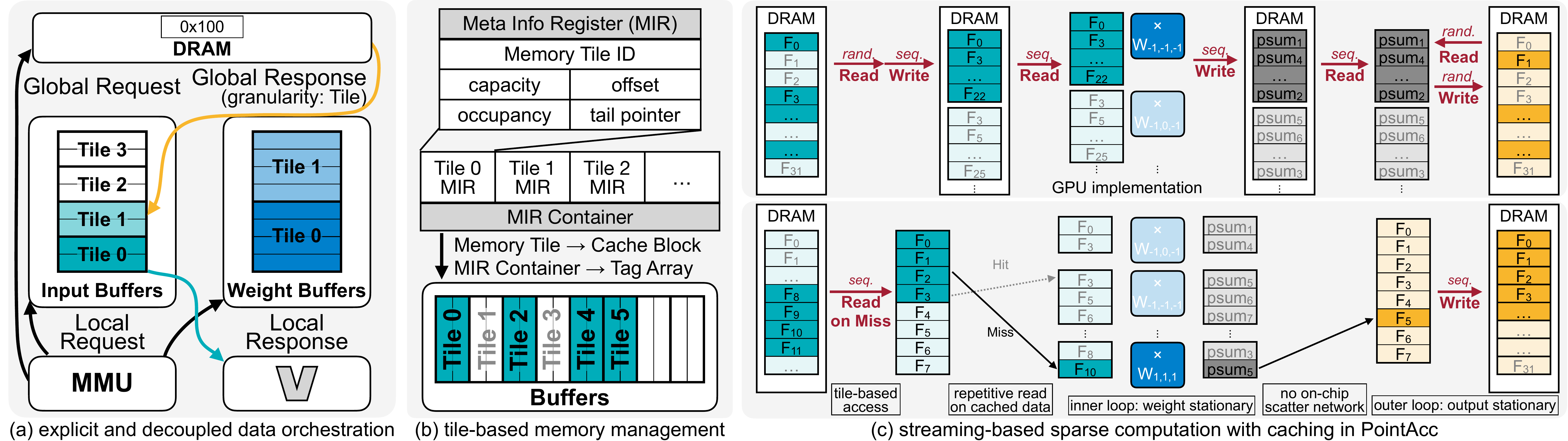}
    \caption{Memory Management Unit (MMU) design overview. (a) \MMU exploits explicit and decoupled data orchestration~\cite{pellauer2019buffets}. (b) \MMU manages the on-chip memory in the granularity of ``tile'' (\ie, block). Its meta information (\eg, allocated capacity, starting address) is recorded in the Memory Tile Meta Info Register (MIR). (c) To handle the sparsity of point cloud convolution, \MMU configures the input buffers as ``cache'' \textit{on demand}, and stream the matrix computation without off-chip scattering.}
    \label{fig:mmu}
\end{figure*}

\subsubsection{\texttt{MergeSort} of Arbitrary Length.}
An $N$-merger can only process a fixed-length array of $N$ elements, typically less than 64, which is far away from the size of point clouds ($10^3$-$10^5$ points). To handle such large scale, we add a forwarding loop after the merger. In each cycle, the merger takes in two arrays of $N/2$ elements but only consumes one array. Only first $N/2$ output elements are considered valid, and the rest $N/2$ elements are buffered for the next cycle. 

\mybfitvparagraph{Example.} \fig{fig:mpu-example}a demonstrates how to achieve the \texttt{MergeSort} of \textit{arbitrary} length. $N=8$ merger merges two input arrays of 4 elements. Thus, we apply a sliding window of size 4 on the input data before feeding to the merger.

At iteration 0, we feed the first 4 elements of both input (yellow) and output (green) point cloud to the merger. Meanwhile, both windows' last elements are compared to determine whose window will be moved forward in the next cycle. Because coordinates (1,1) $<$ (2,0), the window on the output cloud will move forward. Since there could be elements larger than (1,1) but smaller (2,0) in the next cycle, all elements larger than (1,1) in the results should be discarded to ensure correctness. Therefore, (1,4) and (2,0) is marked as invalid by using (1,1) as a threshold. Since it is guaranteed that the merger consumes exactly one window (4 elements) in each iteration, we only output the first 4 elements of the merger results in each cycle. The rest 4 elements will be stored in the register to be used in the next cycle.

At iteration 1,  we update the window of the output cloud (green) and keep that of the input cloud (yellow). As there are 2 valid elements we stored in the register in the last cycle, the first 2 elements of the current merger results are discarded and replaced by the elements from the last cycle. 

\subsubsection{\texttt{Sort}/\texttt{TopK} of Arbitrary Length.}
A single pass of the first 5 stages in \MPU without any forwarding works as a typical bitonic sorter. However, similar to the challenge in \texttt{MergeSort} mentioned above, to handle the \textit{arbitrary} length of inputs, we perform a classical merge sort algorithm by forwarding the outputs of stage MS to stage FM and iteratively merge-sorting two sorted sub-arrays, as shown in \fig{fig:mpu-example}b. By truncating the intermediate subarrays to the length of \textit{k}, \MPU is able to directly support \texttt{TopK} with the same dataflow as running \texttt{Sort} operation as illustrate in \fig{fig:mpu-example}c.

Since the $k$ of \texttt{TopK} in point cloud models is usually very small (\eg, 16/32/64) compared to the size of input (\eg, 8192), the overhead of reusing sorter would be negligible. Experiment shows that on average our design is 1.18$\times$ faster than the quick-selection-based top-k engine proposed in SpAtten~\cite{wang2021spatten} with the same parallelism.

\subsection{\MemoryManagementUnit}

As pointed out in Bottleneck II, explicit gather and scatter hinder the matrix computation. Therefore, we specialize the \MemoryManagementUnit (\MMU) to bridge the gap between computational resource needs and memory bound.

\subsubsection{Data Orchestration}
In point cloud networks, \#layers of sparse computation (point cloud convolution) and dense computation (FC, convolution with kernel size of 1) are comparable. Among traditional memory idioms, workload-agnostic cache design is favoured for sparse computation, while workload-controlled scratchpad is popular for dense computation~\cite{pellauer2019buffets}. To better handle both types of computation, \MMU hybridizes two memory designs. \MMU decouples the memory request initiator and response receiver (\fig{fig:mmu}a), and manages the on-chip buffers in the granularity of ``tile'' (\fig{fig:mmu}b). A memory tile contains the minimum memory space required for a computation tile of tiled matrix multiplication. The memory tile information such as address range and starting offset is exposed in the Memory Meta Info Register (\MIR). Therefore, \MMU is able to perform explicit and precise control over the memory space by manipulating the placement and replacement of MIRs in the \MIRContainer (\fig{fig:mmu}b): \MMU will treat the \MIRContainer as a Tag Array when cache is needed for sparse computation, and as a FIFO or Stack when scratchpad is needed for dense computation.
\begin{figure*}[t]
    \centering
    \includegraphics[width=\linewidth]{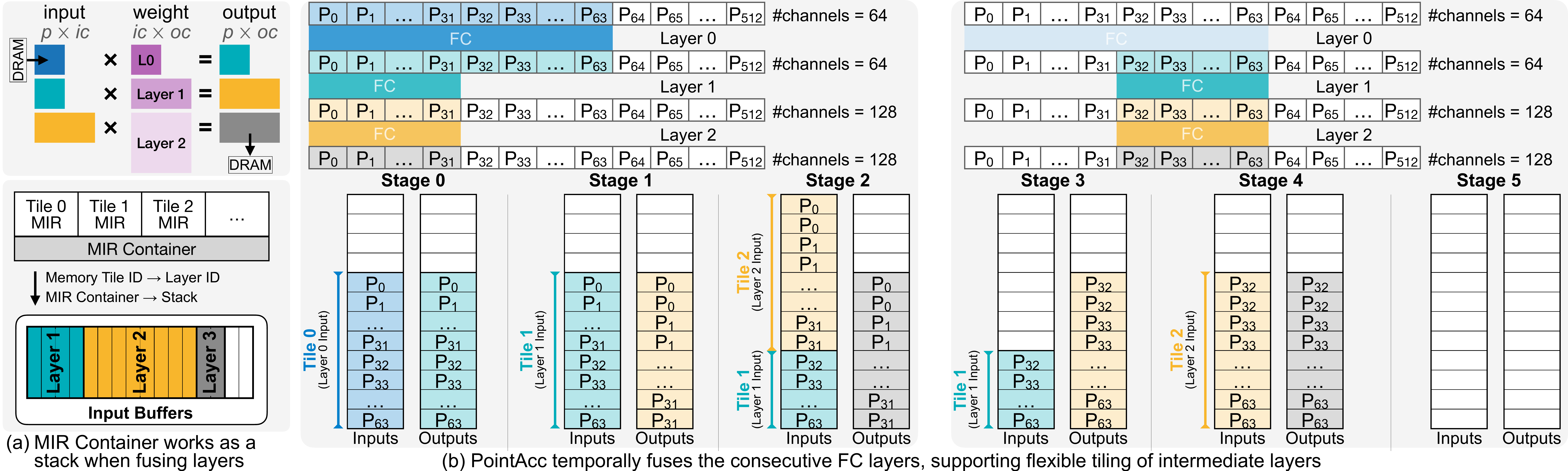}
    \caption{\arch temporally fuses the consecutive FC layers: (a) \MIRContainer is configured as a stack where different MIRs represent the data of different layers. (b) The data of current layer being computed are always at the top of the stack (\eg, Layer 2 in Stage 2). A memory tile is released if all the data are used (\eg, Layer 0 in Stage 1). If there are unused data for the previous layers, MIR only releases the used part (\eg, Layer 1 tile in Stage 2 is halved compared to that in Stage 1).}
    \label{fig:mmu-fuse}
\end{figure*}
\subsubsection{Data Flow}
By reordering the computation loops of matrix multiplication, one can easily realize different data flow to improve the data reuse for inputs/weights/outputs. Since \#points ($10^3\sim 10^5$) is much larger than the \#channels ($10 \sim 10^3$), even by orders of magnitude, we opt \textit{weight stationary} data flow for inner computation loop nests to reduce on-chip memory accesses for weights. \MMU will not increment input/output channels or neighbor dimension until it traverses all points in the on-chip buffers. Furthermore, we opt \textit{output stationary} data flow for outer loop nests to \textit{eliminate} the off-chip scatter of partial sums. \MMU will not swap out the output features before it traverses all neighbors and all input channels. 

\subsubsection{\MMU for Sparse Computation}
\label{sect:mmu-fetch-on-demand}
As discussed in \sect{sect:background}, MatMul computation in point cloud convolution is sparse and guided by the \mappingpairs generated by \MappingUnit. Thus, in addition to computation loop counters, address generator will also require information from the \mappingpairs, as shown in \fig{fig:architecture} (top).

\mybfitvparagraph{Optimize the Computation Flow.} Computation flow affects memory behavior. The \sota GPU implementation will first gather all required input feature vectors, concatenate them as a contiguous matrix and then apply MM to calculate partial sums, referred as Gather-MatMul-Scatter flow. Contrarily, \arch calculates Matrix-Vector multiplication immediately after fetching the input features, referred as \textit{Fetch-on-Demand flow}. As shown in \fig{fig:mmu}c, Fetch-on-Demand flow will save the DRAM access for input features by at least 3\x, by reducing the repetitive reads for gather and eliminating the writes after gather and reads for MM in the Gather-MatMul-Scatter flow.

\mybfitvparagraph{Configure Input Buffers as Cache.} In order to further reduce the repetitive reads of the same input point features in the Fetch-on-Demand flow, MMU configures the Input Buffers as a direct-mapped \textit{cache} by reusing the \MIRContainer as a shared Tag Array recording the feature information.

Different from the traditional cache, contiguous entries in the input buffers are treated as a ``cache block'', and thus the block size is software-controllable. The tag is composed of the point index and channel index of \textit{first} input point inside the cache block.
If both input point index and channel index of requested point features lie in the cache block size, a cache hit occurs. Otherwise, it is a cache miss and MMU will load the data block (\ie, a memory tile) containing the requested point features from DRAM.

\fig{fig:cache-miss-rate} shows the cache miss rate under \streaming mode running \SparseConv of different parameters, where \texttt{k} is the kernel size (\ie, \#neighbors) and \texttt{c} is the \#channels.
Both higher \#neighbors and \#channels improve the input features' chances of being reused, and thus lower the cache miss rate. Meanwhile, as cache block size increases, the cache miss rate decreases as well but saturates at different points for different convolution parameters. Since a larger block size requires a longer latency to load from DRAM (\ie, a larger miss penalty), MMU is configured with different block sizes when running different \SparseConv layers.

\subsubsection{\MMU for Dense Computation}
For FC layers and convolutions with kernel size of 1, the matrix computation is dense and the input features are already contiguous in the DRAM. Thus, \MMU only queries the \MIRContainer at the very beginning of each computation loop tile, reading out the \MIR of current til and trying to allocate the memory and prefetch the data for next computation tile. A memory tile will be evicted (\ie, release) only when it conflicts with the prefetching tile. Hence, data will stay on-chip as long as possible, and be reused as much as possible.

\mybfitvparagraph{Layer Fusion.} \arch is able to fuse the computation of consecutive FCs in the point cloud convolution (\sect{sect:background-matmul-operations}) to further eliminate the DRAM accesses of the intermediate features. The conventional layer fusion~\cite{alwani2016fusedlayer} spatially pipelines the computation, and thus fixes the number of fused layers and requires matching throughput in between.
However, the number of consecutive layers varies among different models and even among different blocks in the same model. \MMU thus \textit{temporally} fuses these computation by simply configuring the \MIR Container works as \textit{Stack} and identifying the \MIR by the layer index. The \MatrixComputingUnit always works with the top entry of \MIR Container. When switching back to the previous unfinished layer, \MIR Container will pop the top entry.

For FCs in the point cloud models, the point dimension can be regarded as the batch size dimension in traditional CNN. Thus, \MMU can simplify the layer fusion logic by tiling the point dimension only without any halo elements between tiles. 
The number of fused layers and their tilings are determined during the compilation to avoid memory overflow. For each set of consecutive FCs, we will try to fuse all unprocessed FCs. If the estimated memory of required intermediate data overflows for all possible tilings, we will discard the last layer and try to fuse remaining ones. Such process is repeated until all layers are processed.

\mybfitvparagraph{Example.} \fig{fig:mmu-fuse}b shows an example of \arch fusing 3 consecutive FC layers.
\begin{itemize}[topsep=0pt, leftmargin=18pt]
    \item \textbf{Stage 0}: \MMU loads features of $p_0$ to $p_{63}$ of layer 0 from DRAM.
    \item \textbf{Stage 1}: the computation in Stage 0 used up all loaded data and thus layer 0 tile is released. Switching to layer 1, \MMU pushes features of layer 1 from Output Buffers. 
    \item \textbf{Stage 2}: since layer 1 computation only uses half of input features ($p_0$ to $p_{31}$), the layer 1 tile capacity is halved. Switching to layer 2, \MMU pushes features of layer 2 similar to Stage 1.
    \item \textbf{Stage 3}: since layer 2 is the last fused layer, we switch back to the previous layer (layer 1) after \MMU pops layer 2 data. Since layer 1 tile capacity is nonzero, we continue to compute layer 1 for the rest features ($p_{32}$ to $p_{63}$).
    \item \textbf{Stage 4}: layer 1 tile is released since all data are used. Switching to layer 2, \MMU pushes features of layer 2 similar to Stage 2.
    \item \textbf{Stage 5}: after finishing layer 2 and switching back to layer 1, we find that layer 1 tile capacity is zero. Thus we continue switching back to layer 0. Since layer 0 tile capacity is also zero, we will update outer loop counters, and then continue to work on layer 0 for the next tile $p_{64}$ to $p_{127}$.
\end{itemize}

\subsection{\MatrixComputingUnit}
\MatrixComputingUnit (\MCU) adopts the classic systolic array as the computing core, since it has been proven to be efficient and effective for dense matrix multiplication. 
In order to completely eliminate the scatter network on-chip, \MCU parallelizes the computation in input channels (\texttt{ic}) and output channels (\texttt{oc}) dimensions: each row of PEs computes in \texttt{ic} and each column of PEs computes in \texttt{oc} dimension independently. Thus, \MCU only accesses features of one output point in one cycle, no more spatially scattering different points at one time and thus no need for the scatter circuit.

\begin{table}[t]
  \setlength{\tabcolsep}{1pt}
  \centering
  \caption{Evaluation Benchmarks}
  \scalebox{0.8}{
    \begin{tabular}{ccccc}
    \toprule
    \textbf{Application} & \textbf{Dataset} & \textbf{Scene} & \textbf{Model} & \textbf{Notation} \\
    \midrule
    \multirow{2}[2]{*}{Classification} & \multirow{2}[2]{*}{ModelNet40~\cite{wu2015modelnet}} & \multirow{2}[2]{*}{Object} & PointNet~\cite{qi2017pointnet} & PointNet \\
          &       &       & PointNet++ (SSG)~\cite{qi2017pointnet++} & PointNet++(c)  \\
    \midrule
    Part  & \multirow{2}[2]{*}{ShapeNet~\cite{chang2015shapenet}} & \multirow{2}[2]{*}{Object} & PointNet++ (MSG)~\cite{qi2017pointnet++} & PointNet++(ps) \\
    Segemantation &       &       & DGCNN~\cite{wang2018dgcnn} & DGCNN \\
    \midrule
    Detection & KITTI~\cite{geiger2012kitti} & Outdoor & F-PoinNet++~\cite{qi2017frustum} & F-PointNet++ \\
    \midrule
    \multirow{3}[3]{*}{Segemantation} & \multirow{2}[1]{*}{S3DIS~\cite{armeni2016s3dis}} & \multirow{2}[1]{*}{Indoor} & PointNet++ (SSG)~\cite{qi2017pointnet++} & PointNet++(s) \\
          &       &       & MinkowskiUNet~\cite{choy20194d} & MinkNet(i) \\
\cmidrule{2-5}
    & SemanticKITTI~\cite{behley2019semantickitti} & Outdoor & MinkowskiUNet~\cite{choy20194d} & MinkNet(o) \\
    \bottomrule
    \end{tabular}%
}
  \label{tab:benchmarks}
\end{table}
\begin{table}[t]
\setlength{\tabcolsep}{5pt}
  \centering
  \renewcommand{\arraystretch}{0.85}
  \caption{Evaluated ASIC platforms}
  \scalebox{0.9}{
    \begin{tabular}{l|ccc}
    \toprule
    \textbf{Chip} & \textbf{Mesorasi} & \textbf{\arch} & \textbf{\archsmall} \\
    \midrule
    \textbf{Cores} & 16$\times$16=256 & 64$\times$64=4096 & 16$\times$16=256 \\
    \midrule
    \textbf{SRAM (KB)} & 1624  & 776   & 274 \\
    \midrule
    \textbf{Area (mm2)} & -     & 15.7  & 3.9 \\
    \midrule
    \textbf{Frequency} & 1GHz  & 1 GHz & 1 GHz \\
    \midrule
    \textbf{DRAM} & LPDDR3-1600 & HBM 2 & DDR4-2133 \\
    \textbf{Bandwidth} & 12.8 GB/s & 256 GB/s & 17 GB/s \\
    \midrule
    \textbf{Technology} & 16 nm & 40 nm & 40 nm \\
    \midrule
    \textbf{Peak Performance} & 512 GOPS & 8 TOPS & 512 GOPS \\
    \bottomrule
    \end{tabular}%
    }
  \label{tab:asics}%
\end{table}%
\begin{figure*}[t]
\begin{minipage}[t]{0.49\linewidth}
    \centering
    \includegraphics[width=\linewidth]{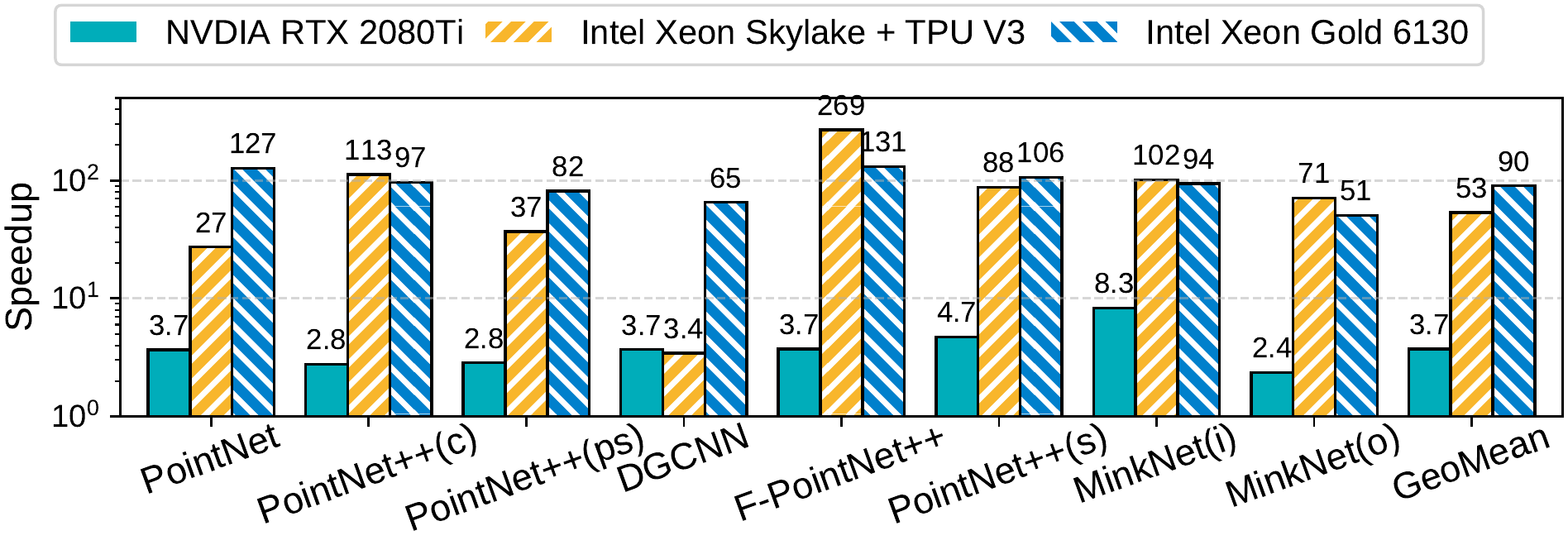}
    \includegraphics[width=\linewidth]{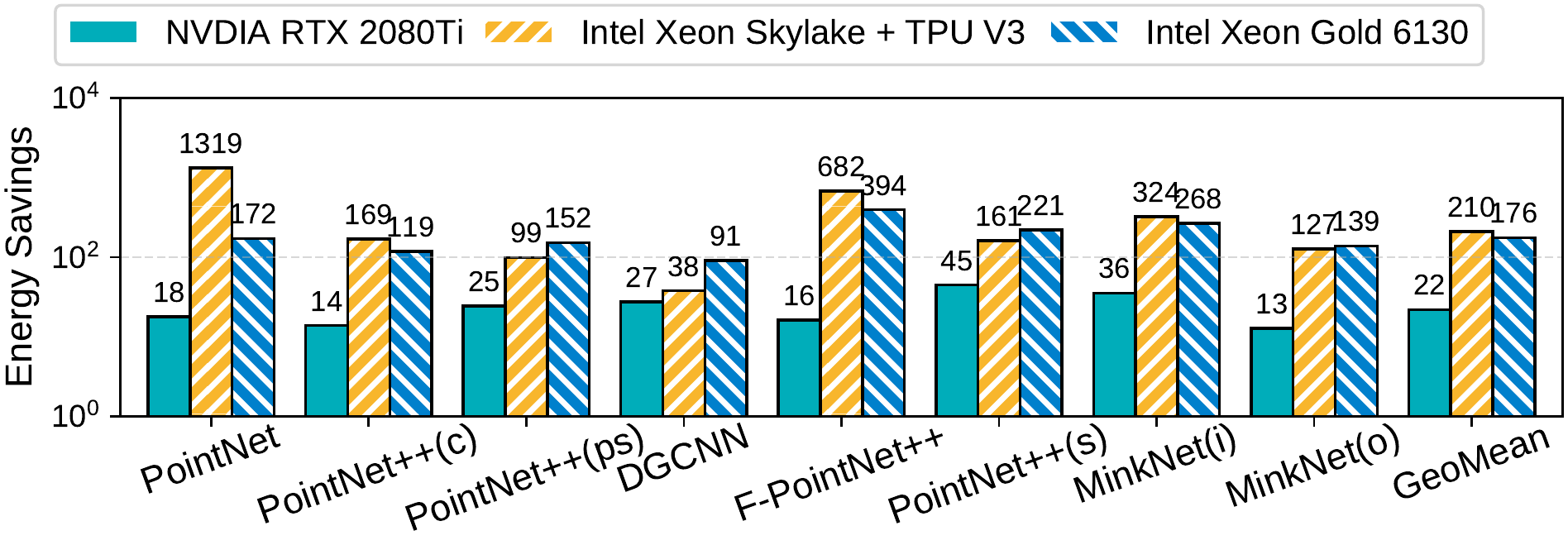}
    \caption{Performance gain over the server products: \arch is 3.7\x faster than RTX 2080Ti on average.}
    \label{fig:performance-large}
\end{minipage}
\begin{minipage}[t]{0.49\linewidth}
    \centering
    \includegraphics[width=\linewidth]{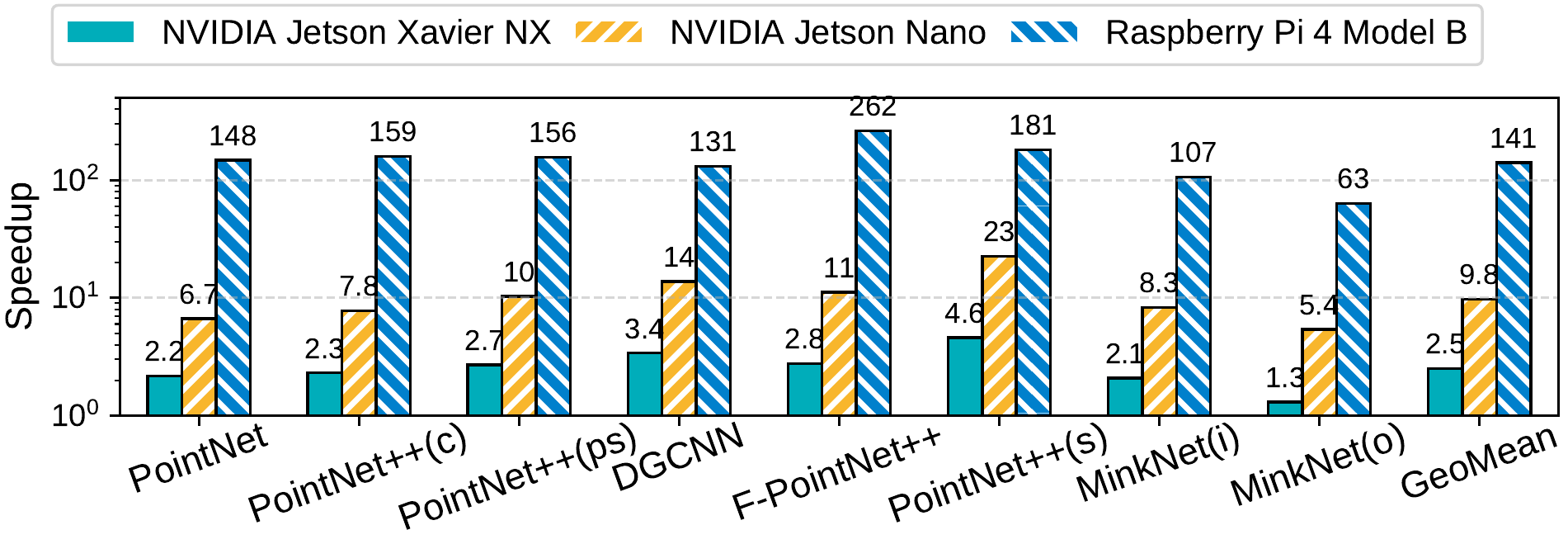}
    \includegraphics[width=\linewidth]{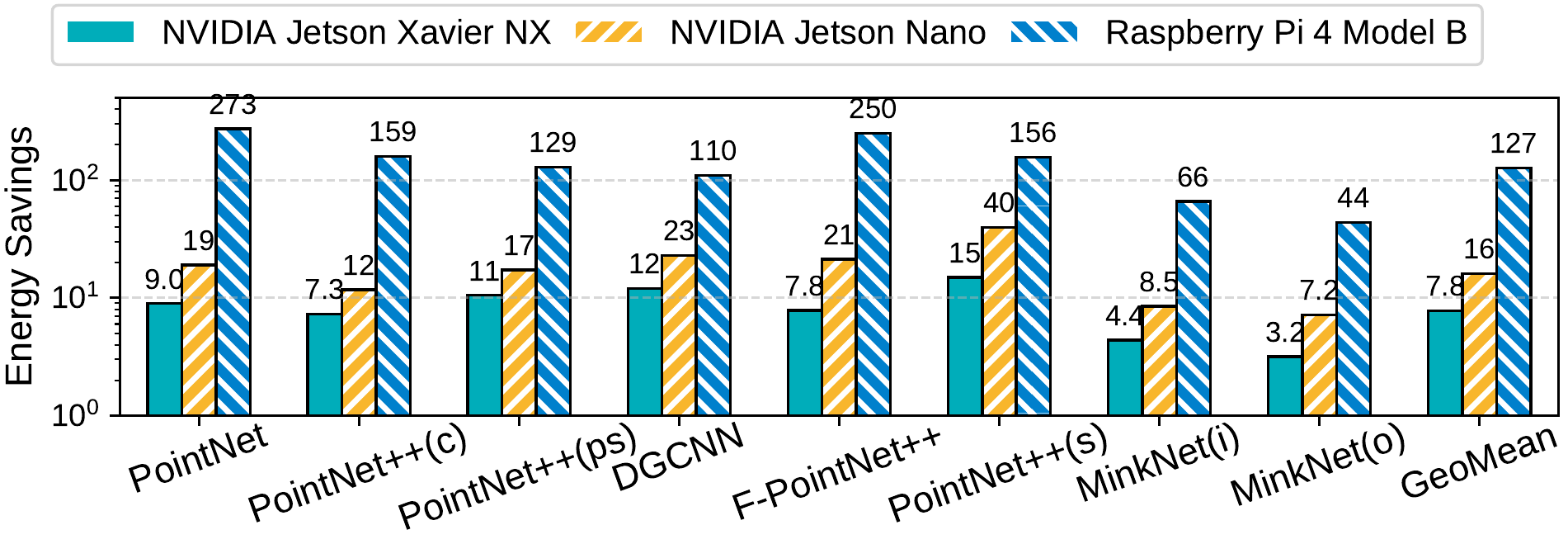}
    \caption{Performance gain over the edge devices: \archsmall is 2.5\x faster than Jetson NX on average.}
    \label{fig:performance-small}
\end{minipage}
\end{figure*}
\section{Evaluation}
\subsection{Evaluation Setup}
\bfparagraph{Benchmarks.}
We pick 8 diverse point cloud network benchmarks across assorted application domains, including object classification, part segmentation, detection, and semantic segmentation. As shown in \tab{tab:benchmarks}, the networks include both classical and state-of-the-art ones and cover all categories in \tab{tab:block-structure}. The selected datasets also contain various sizes and modalities for input point clouds, from daily objects to indoor scenes to spacious outdoor scenes. Such extensive benchmarks allow us to evaluate \arch thoroughly.

\bfparagraph{Hardware Implementation.}
We implement \arch with Verilog and verify the design through RTL simulations. We synthesize \arch with Cadence Genus under TSMC 40nm technology. Power of \arch is simulated with fully annotated switching activity generated with the selected benchmarks.
We develop a cycle-accurate simulator to model the exact behavior of the hardware and calculate the cycle counts and read/write of on-chip SRAMs. The simulator is also verified against the Verilog implementation. We integrate the simulator with Ramulator~\cite{kim2015ramulator} to model the DRAM behaviors. We obtain SRAM energy with CACTI~\cite{cacti}, and DRAM energy using Ramulator-dumped DRAM commands trace.

\bfparagraph{Baselines.}
We adopt three kinds of hardware platforms as evaluation baselines: server-level products, edge devices, and a specialized ASIC design. For server-level products, we compare full-size \arch against 
Xeon\textsuperscript \textregistered\  6130 CPU, RTX 2080Ti GPU, and TPU-v3. For edge devices and specialized ASIC, we compare the edge configuration (\archsmall) against Jetson Xavier NX, Jetson Nano, Raspberry Pi 4B, and Mesorasi~\cite{feng2020mesorasi} with NPU of 16$\times$16 systolic array.
ASIC design parameters are compared in \tab{tab:asics}.

We implement point cloud networks with PyTorch (matrix operations with Intel MKLDNN / cuDNN, and \sparsemapping operations with C++ OpenMP / custom CUDA kernels). Our implementation achieves 2.7$\times$ speedup over the \sota implementaion \texttt{MinkowskiEngine}~\cite{choy20194d}.

\begin{figure*}[t]
\begin{minipage}[b]{0.485\linewidth}
    \centering
    \includegraphics[width=\linewidth]{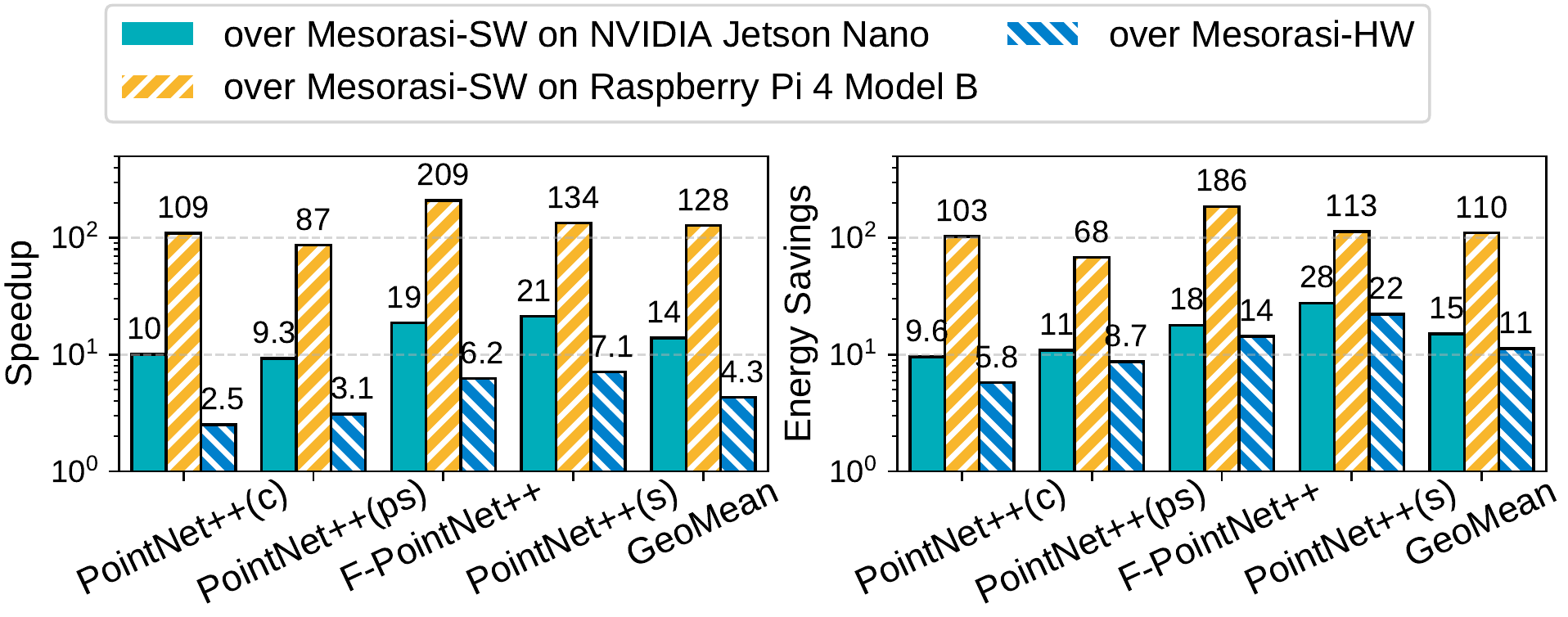}
    \vspace{-15pt}
    \caption{Speedup and energy savings of \archsmall over Mesorasi. Mesorasi-SW runs Mesorasi networks without specialized architectural support; Mesorasi-HW executes Mesorasi networks with dedicated aggregation unit AU and neural processing unit NPU.}
    \label{fig:mesorasi}
\end{minipage}
\hfill
\begin{minipage}[b]{0.485\linewidth}
    \centering
    \includegraphics[width=0.9\linewidth]{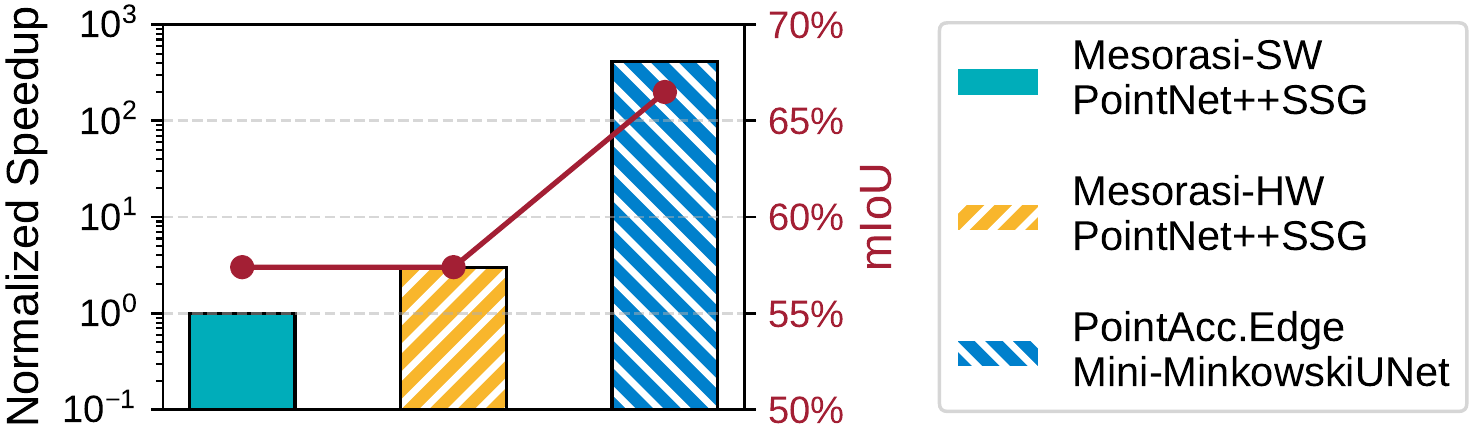}
    \caption{Mesorasi does not support independent weights for different neighbors, which is crucial in some variants of \pointnetpp-based blocks~\cite{qi2017pointnet++} or \SparseConv-based blocks~\cite{choy20194d}. When running the same segmentation task on S3DIS dataset, \archsmall is able to execute networks with 10\% higher accuracy and 100\x lower latency.}
    \label{fig:mesorasi-s3dis}
\end{minipage}
\end{figure*}
\begin{figure}[t]
\centering
\includegraphics[width=\linewidth]{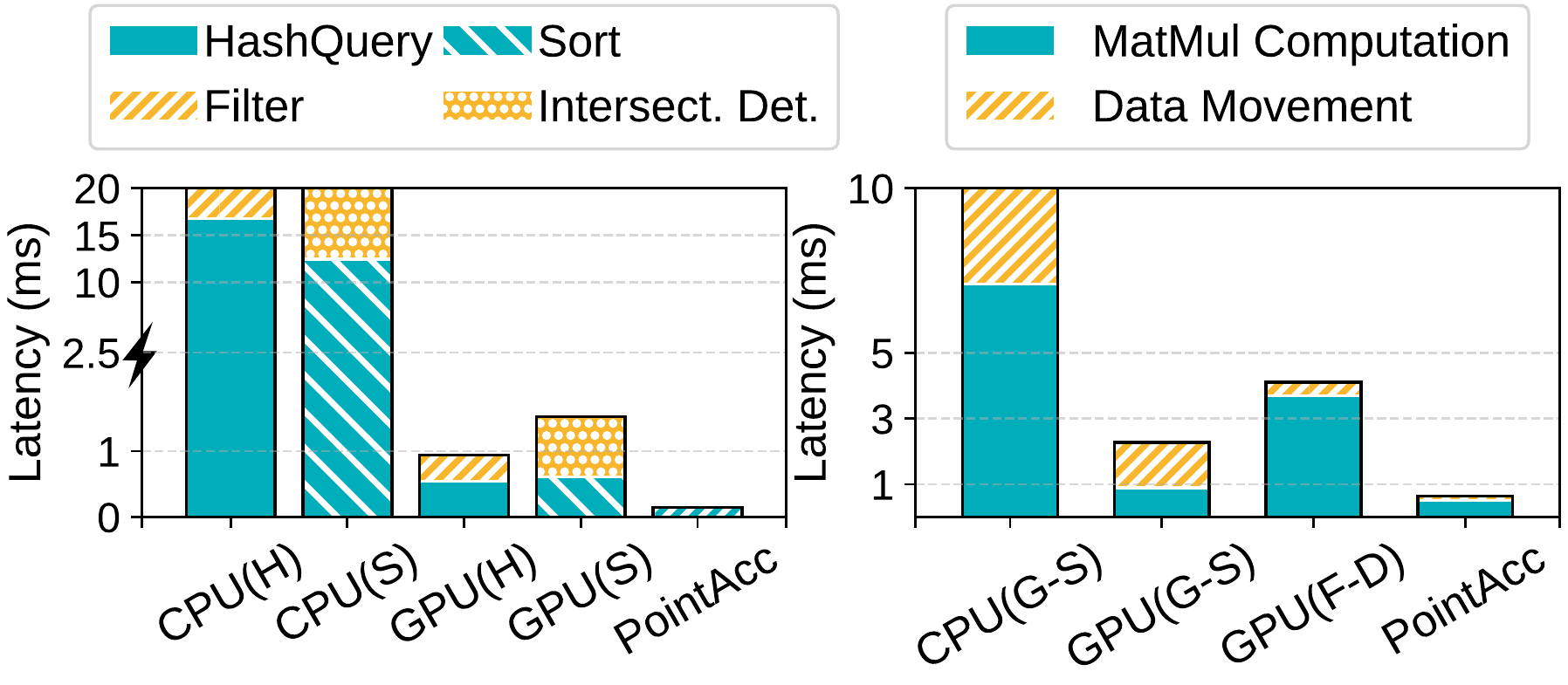}
\caption{Latency breakdown of \SparseConv-based blocks in operation level. 
Left: Kernel Mapping. Mergesort-based solution (\textit{S}) runs slower than hash-table-based algorithm (\textit{H}) on CPU/GPU, but 1.4\x faster with 14\x smaller area after circuit specialization (\sect{sect:unify-mapping-ops}).
Right: Convolution. Compared to the Gather-MatMul-Scatter flow (\textit{G-S}), the reduction of data movement in \Streaming flow (\textit{F-D}) is dwarfed by the MV computation cost on GPU, but benefits \arch (\sect{sect:mmu-fetch-on-demand}).}
\label{fig:bottleneck-operation}
\end{figure}

\subsection{Evaluation Results}
\subsubsection{Speedup and Energy Savings} 
\fig{fig:performance-large} presents the performance and energy benefits of \arch in comparison with GPU, TPU, and CPU products. On average, \arch offers 3.7\x, 53\x, 90\x speedup and 22\x, 210\x, 176\x energy savings, respectively. \fig{fig:performance-small} shows the speedup and energy savings of \archsmall over Jetson Xavier NX, Jetson Nano, and Raspberry Pi devices. On average, \archsmall achieves 2.5\x, 9.8\x, 141\x speedup, and 7.8\x, 16\x, 127\x energy savings, respectively. The improvements are \textit{consistent} on different benchmarks. For TPU V3, the considerable gain mainly comes from supporting \sparsemapping operations with share compute kernel design, thus significantly reducing the data movements to/from host CPU.

\subsubsection{Comparison with \mesorasi Architecture}
\label{sect:eval-mesorasi}
\fig{fig:mesorasi} shows the runtime speedup and energy savings of \archsmall over \mesorasi designs on \pointnetpp-based benchmarks. \archsmall achieves 1.3\x speedup and 11\x energy savings over \mesorasi hardware design on average.
Unlike our design, \mesorasi is limited since it does not support independent weights for different neighbors. It is crucial for many point cloud networks~\cite{li2018pointcnn, choy20194d, tang2020searching, wu2019pointconv}, especially for \SparseConv-based models which not only improve the accuracy but also are capable of processing large-scale point clouds. We compare \archsmall with \mesorasi for the same segmentation task on the indoor scene dataset S3DIS. We scale down the \SparseConv-based state-of-the-art model MinkowskiUNet to a shallower, narrower version, denoted as Mini-MinkowskiUNet. Co-designed with the neural network, \archsmall delivers over 100\x speedup and improves the mean Intersection-over-Unit (mIoU) accuracy by \mesorasiAcc.

\subsubsection{Source of Performance Gain}

\bfparagraph{Ranking-based conversion of \sparsemapping operations.} \fig{fig:bottleneck-operation} (left) breaks down the latency of mapping operation (\kernelmapping here) of the 1st downsampling \SparseConv-based block on SemanticKITTI. The mergesort-based algorithm (\fig{fig:spconv-example-alg}) even worsens the performance on CPU/GPU. But compared to CPU/GPU, \arch is over 10\x faster. Our design provides enough parallelism for comparison-intensive merge sort, and eliminates the repetitive access on intermediate results by spatially pipelining the stages of merge sort and intersection detection. On CPU/GPU, detecting intersection costs almost 2\x runtime than filtering the query miss on hash table, because the length of inputs of intersection detection is doubled due to the merge of input/output point clouds. Querying the hash table costs almost comparable time with merge sort on GPU. It is because the hash-table-based algorithm only needs one pass on the input, but most stages of bitonic merge require a scan of the input from GPU global memory.

\bfparagraph{\Streaming computation flow.} \fig{fig:bottleneck-operation} (right) breaks down the latency of the matrix multiplication (convolution here) of the 1st layer of MinkowskiUNet on SemanticKITTI. \Streaming flow saves the memory footprint by 3\x but decomposes the matrix-matrix multiplication into fragmentary matrix-vector multiplications, and thus significantly increases the overhead due to low utilization of GPU. However, such overhead is removed in \arch because of the computation power of the systolic array. Therefore, \arch spends almost comparable time on the whole operation against the MatMul computation part only in the Gather-MatMul-Scatter flow.

\begin{figure*}[t]
\begin{minipage}[t]{0.325\linewidth}
    \centering
    \includegraphics[width=0.8\linewidth]{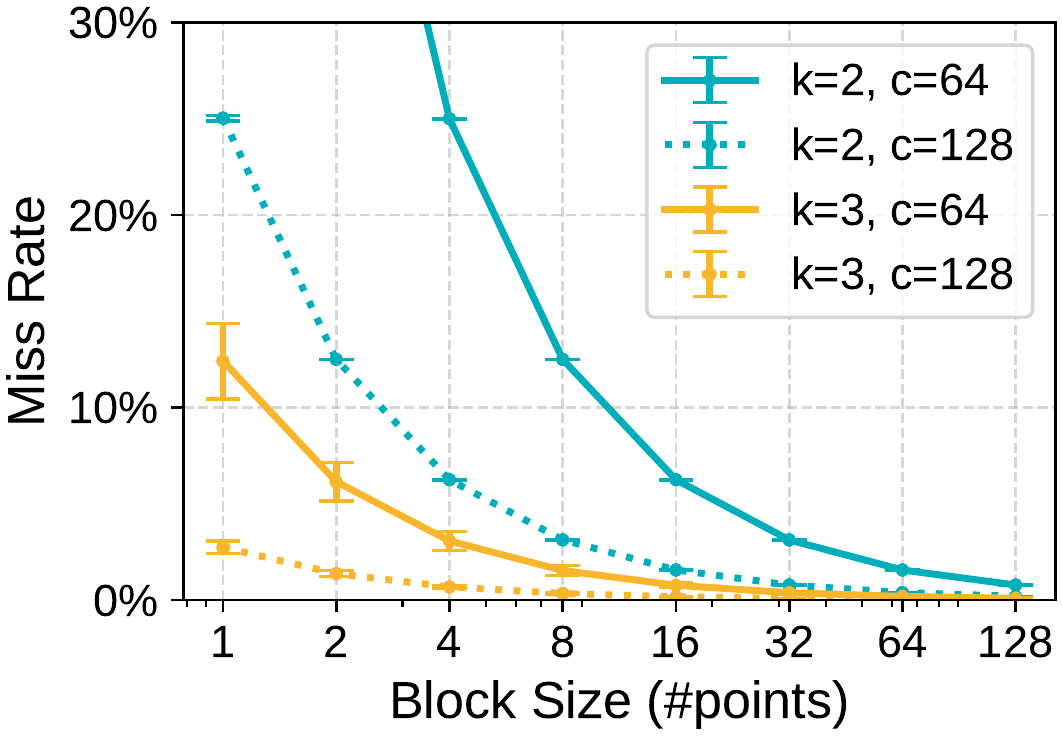}
    \caption{Miss rate of cache mode memory management. Miss rate decreases as cache block size, kernel size and \#out\_channels increases.}
    \label{fig:cache-miss-rate}
\end{minipage}
\hfill
\begin{minipage}[t]{0.325\linewidth}
    \centering
    \includegraphics[width=0.9\linewidth]{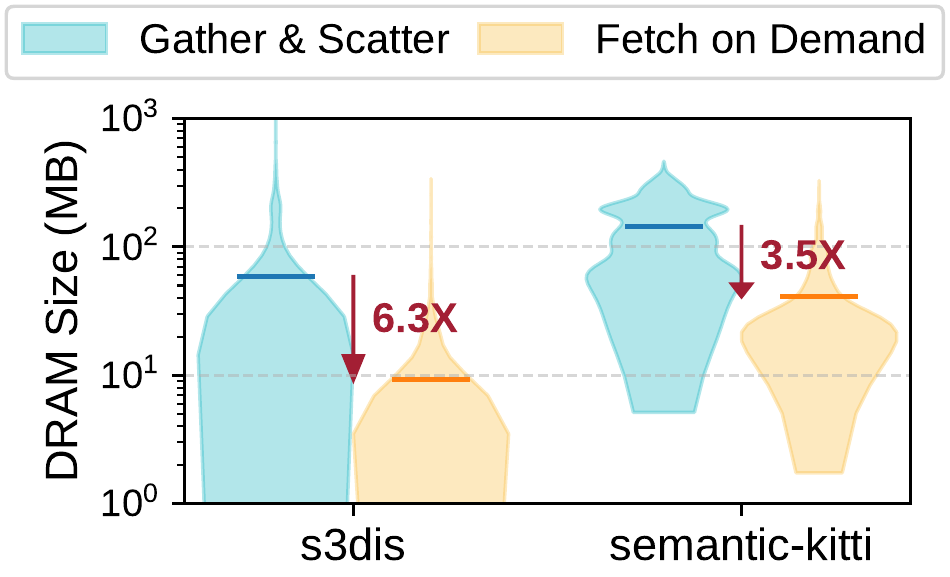}
    \caption{MinkowskiUNet layer DRAM access size distribution with and without caching. The bar denote the average DRAM access size per layer.}
    \label{fig:dram-reduction-cache}
\end{minipage}
\hfill
\begin{minipage}[t]{0.325\linewidth}
    \centering
    \includegraphics[width=0.85\linewidth]{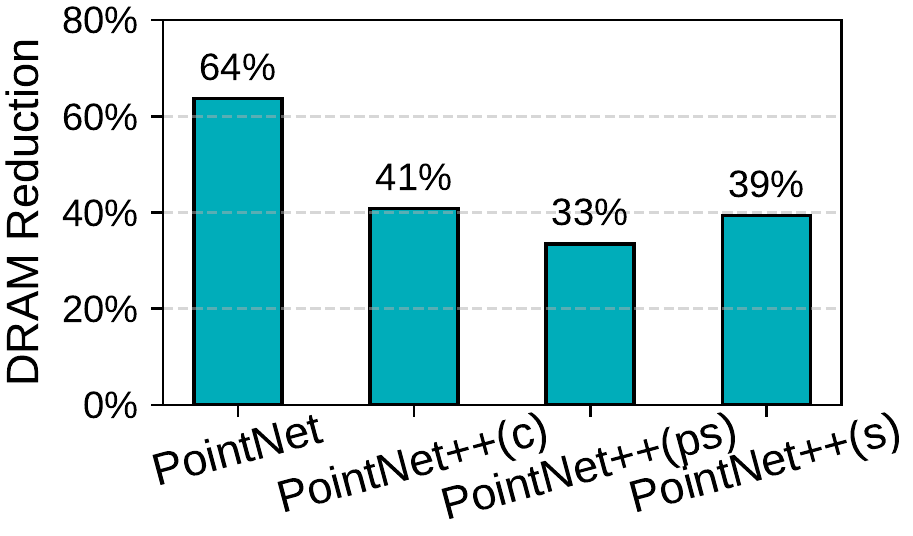}
    \caption{PointNet++ DRAM access size reduction of fusion mode memory management compared to running layer by layer independently.}
    \label{fig:dram-reduction-fuse}
\end{minipage}
\end{figure*}
\begin{figure}[t]
    \hfill
    \begin{subfigure}[t]{0.45\linewidth}
    \centering
    \includegraphics[width=\linewidth]{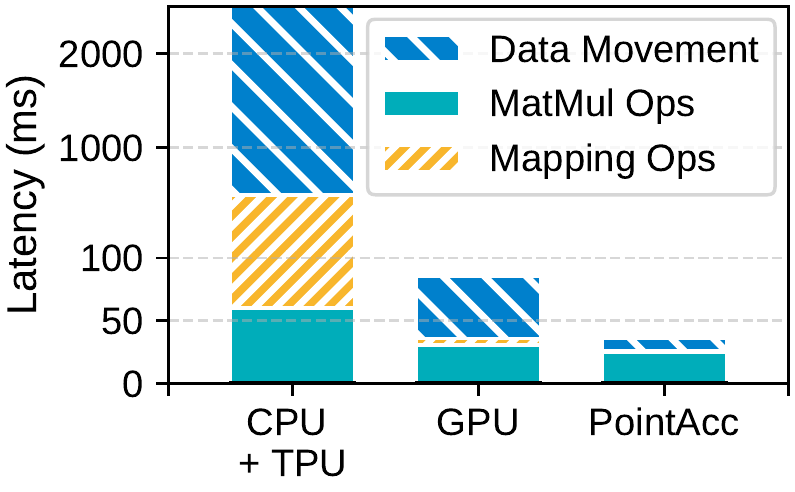}
    \caption{Latency breakdown}
    \label{fig:latency-breakdown}
    \end{subfigure}
    \hspace{5pt}
    \begin{subfigure}[t]{0.45\linewidth}
    \centering
    \includegraphics[width=0.825\linewidth]{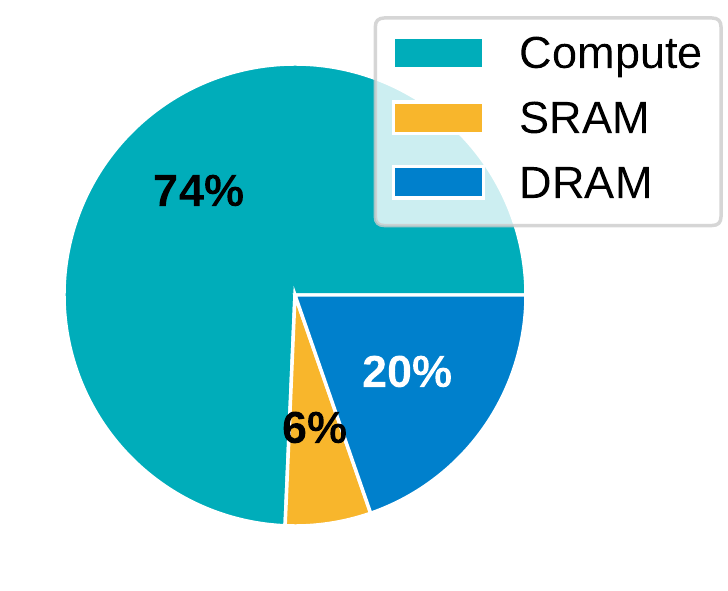}
    \caption{Energy breakdown}
    \end{subfigure}
    \hfill
    \caption{Performance breakdown of \arch on MinkNet(o) benchmark: \arch reduces the latency and energy cost of data movement.}
    \label{fig:energy-breakdown}
\end{figure}

\bfparagraph{Configurable caching.} \fig{fig:dram-reduction-cache} demonstrates the distribution (\ie, probability density) of the DRAM access size per layer in MinkowskiUNet on S3DIS and SemanticKITTI dataset. A wider region indicates higher frequency of the given data size. The shape of distribution are nearly the same with/without caching, which indicates that the caching works \textit{consistently} on different layers and on different datasets. On average, the configurable cache reduces the layer DRAM access by 3.5\x to 6.3\x, where each point features are only fetched nearly once on average. 

\bfparagraph{Temporal layer fusion.} \fig{fig:dram-reduction-fuse} shows the reduction ratio of DRAM access when running \pointnetpp-based networks with Fusion Mode. Compared against running networks layer by layer independently, our simplified layer fusion logic help cut the DRAM access from 33\% to 64\%. Since there is no downsampling layers in PointNet, we are able to fuse more layers than other \pointnetpp-based networks, which leads to 1.5\x to 2\x more DRAM reduction.

\bfparagraph{Overall performance breakdown.} \fig{fig:energy-breakdown} breaks down the latency and energy consumption of \arch running MinkowskiUNet on SemanticKITTI dataset. The support of \sparsemapping operations gets rid of the extra cost on communication between co-processors as in CPU+TPU case. The specialized data orchestration in \MMU helps reduce the DRAM access and make the data movement overlap the matrix multiplication. Therefore, the MatMul operations dominate the overall latency. Moreover, the computation covers 69\% of total energy while DRAM access costs 23\% of total energy, which differs from the observation in~\cite{chen2016eyeriss} where DRAM accesses dominate the energy consumption of FC layers.

\section{Related Work}
\bfparagraph{Deep Learning on Point Clouds.}
Early research converted point clouds to the volumetric representation and applied vanilla 3D CNNs~\cite{qi2016volumetric} on the voxel grids. Qi~\etal then proposed PointNet~\cite{qi2017pointnet} and its hierarchical version, PointNet++~\cite{qi2017pointnet++} to perform direct deep learning on point clouds. Later research such as Deep KD-Net~\cite{klokov2017escape}, SpiderCNN~\cite{xu2018spidercnn}, PointCNN~\cite{li2018pointcnn}, PointConv~\cite{wu2019pointconv}, KPConv~\cite{thomas2019kpconv} are variants of PointNet++. PVCNN~\cite{liu2019pvcnn} combined the advantages of point cloud and volumetric representations. Another stream of research SSCN~\cite{graham20183d}, MinkowskiNet~\cite{choy20194d}, SPVNAS~\cite{tang2020searching} focus on SparseConv-based methods, which are more efficient than PointNet++-based methods on large outdoor scenes.


\bfparagraph{Point Cloud Accelerators.} Researchers have extensively developed architectures and systems~\cite{gieseke2014buffer, qiu2009gpu, heinzle2008hardware, winterstein2013fpga, xu2019tigris} for accelerating neighbor search, especially for point cloud registration task. There has been limited work on point cloud deep learning. Feng \etal proposed Mesorasi, an architecture support for \pointnetpp-based networks via delayed-aggregation~\cite{feng2020mesorasi}. However, Mesorasi has limited applicability as explained in \sect{sect:eval-mesorasi}.

\bfparagraph{Deep Learning Accelerators.}
Various work has explored both FPGA~\cite{umuroglu2017finn} and ASIC~\cite{chen2014dadiannao, chen2016eyeriss, shafiee2016isaac, zhang2020sparch, wang2021spatten} accelerator architectures for efficient DNN inference. EIE~\cite{han2016eie}, Cambricon-X~\cite{zhang2016cambricon},  Cnvlutin~\cite{albericio2016cnvlutin} and SCNN~\cite{parashar2017scnn} exploited the sparsity in DNNs and speeded up the inference by skipping unstructured zeros. As discussed in \sect{sect:motivation}, these conventional sparse accelerators do not support modern point cloud networks.

\section{Conclusion}
Moving from 2D images, machines start to perceive the world through 3D point clouds to recognize the world better. The rapid development of point cloud deep learning brings new challenges and exciting opportunities for intelligent hardware design. This work presents a specialized point cloud deep learning accelerator \arch, a new advancement that could help bring powerful point cloud AI to real-world applications, from augmented reality on iPhones to autonomous driving of intelligent vehicles, supporting real-time interactions with humans. \arch supports the newly introduced \sparsemapping operations by unifying and mapping them onto a shared ranking-based compute kernel. At the same time, \arch addresses the massive memory footprint problem due to the sparsity of point clouds by streaming the sparse computation with on-demand caching  and temporally fusing the consecutive layers of dense computation. Extensive evaluation experiments show that \arch delivers significant speedup and energy reduction over CPU, GPU, and TPU. \arch paves the way for efficient point cloud recognition.

\begin{acks}
This work was supported by National Science Foundation, Hyundai, Qualcomm and MIT-IBM Watson AI Lab. We also thank AWS Machine Learning Research Awards for the computational resource.
\end{acks}

\bibliographystyle{ACM-Reference-Format}
\bibliography{refs}

\end{document}